\documentclass[11pt,a4paper,DIV=11,numbers=noenddot]{scrartcl}
\usepackage[utf8]{inputenc}
\makeatletter
\DeclareOldFontCommand{\rm}{\normalfont\rmfamily}{\mathrm}
\DeclareOldFontCommand{\sf}{\normalfont\sffamily}{\mathsf}
\DeclareOldFontCommand{\tt}{\normalfont\ttfamily}{\mathtt}
\DeclareOldFontCommand{\bf}{\normalfont\bfseries}{\mathbf}
\DeclareOldFontCommand{\it}{\normalfont\itshape}{\mathit}
\DeclareOldFontCommand{\sl}{\normalfont\slshape}{\@nomath\sl}
\DeclareOldFontCommand{\sc}{\normalfont\scshape}{\@nomath\sc}
\makeatother
\usepackage{amsmath,amssymb,graphicx,scalefnt,enumitem}
\usepackage[absolute]{textpos}
\usepackage[noadjust]{cite}
\usepackage[affil-it]{authblk}
\usepackage{xcolor}
\usepackage{tabularx,booktabs}
\usepackage{xspace}
\usepackage{listings}
\usepackage{algorithm}
\usepackage{pifont}
\usepackage{algorithmicx}
\usepackage{algpseudocode}
\usepackage[final]{showkeys}
\usepackage{filemod}
\usepackage{bold-extra}
\usepackage[font=small,labelfont=bf,format=plain,margin=0.05\textwidth]{caption}
\usepackage{subfig}
\usepackage[parfill]{parskip}
\usepackage{xurl}

\usepackage[pdftitle={Interpolation of Dense and Sparse Rational Functions and other Improvements in FireFly},
  pdfauthor={Jonas Klappert, Sven Yannick Klein, and Fabian Lange},
  pdfkeywords={Finite Field, functional reconstruction, functional interpolation, FireFly},
  bookmarks=true, linktocpage,
  colorlinks=true, allbordercolors=white, allcolors=blue]{hyperref}
\newcommand{\pheaderline}{{\footnotesize TTK-20-07\\P3H-20-010}}
\newcounter{notecount}

\newcommand{\citere}[1]{Ref.\,\cite{#1}}
\newcommand{\citeres}[1]{Refs.\,\cite{#1}}
\newcommand{\code}[1]{\texttt{#1}}
\newcommand{\abbrev}[1]{{\scalefont{1}#1}}
\newcommand{\alg}[1]{Alg.\,(\ref{#1})}
\newcommand{\eqn}[1]{Eq.\,(\ref{#1})}
\newcommand{\eqs}[1]{Eqs.\,(\ref{#1})}
\newcommand{\fig}[1]{Fig.\,\ref{#1}}

\newcommand{\tab}[1]{Tab.\,\ref{#1}}

\newcommand{\sct}[1]{Sect.\,\ref{#1}}
\newcommand{\app}[1]{App.\,\ref{#1}}

\newcommand{\zp}[1]{{\mathbb{Z}}_{#1}}

\def\cpc#1#2#3#4{{\it Comp.~Phys.~Commun.~}\jref{\bf #1}{#2}{#3}, \url{https://dx.doi.org/#4}}
\def\chpc#1#2#3#4{{\it Chin.~Phys.~}\jref{\bf C #1}{#2}{#3}, \url{https://dx.doi.org/#4}}

\def\ijmpa#1#2#3#4{{\it Int.~J.~Mod.~Phys.~}\jref{\bf A #1}{#2}{#3}, \url{https://dx.doi.org/#4}}

\def\jacm#1#2#3#4{{\it J.~ACM~}\jref{\bf #1}{#2}{#3}, \url{https://dx.doi.org/#4}}

\def\jhep#1#2#3#4{{\it JHEP~}\jref{\bf #1}{#2}{#3}, \url{https://dx.doi.org/#4}}

\def\lait#1#2#3#4{{\it Linear~Algebra~Its~Appl.~}\jref{\bf #1}{#2}{#3}, \url{https://dx.doi.org/#4}}

\def\npb#1#2#3#4{{\it Nucl.~Phys.~}\jref{\bf B #1}{#2}{#3}, \url{https://dx.doi.org/#4}}
\def\npps#1#2#3#4{{\it Nucl.~Phys.~B~Proc.~Suppl.~}\jref{\bf #1}{#2}{#3}, \url{https://dx.doi.org/#4}}
\def\plb#1#2#3#4{{\it Phys.~Lett.~}\jref{\bf B #1}{#2}{#3}, \url{https://dx.doi.org/#4}}

\def\prd#1#2#3#4{{\it Phys.~Rev.~}\jref{\bf D #1}{#2}{#3}, \url{https://dx.doi.org/#4}}

\def\tit#1#2#3#4{{\it IEEE~Trans.~Inf.~Theory~}\jref{\bf #1}{#2}{#3}, \url{https://dx.doi.org/#4}}

\def\apvmmpi#1#2#3#4{{\it Adv.\ Parallel Virtual Machine Message Passing Interface.\ EuroPVM/MPI~}\jref{\bf #1}{#2}{#3}, \url{https://dx.doi.org/#4}}
\def\esam#1#2#3#4{{\it Symbolic Algebraic Comp.\ EUROSAM~}\jref{\bf #1}{#2}{#3}, \url{https://dx.doi.org/#4}}
\def\focs#1#2#3#4{{\it Proc.\ Symp.\ Foundations Comp.\ Sci.~}\jref{\bf #1}{#2}{#3}, \url{https://dx.doi.org/#4}}
\def\issacfirst#1#2#3#4{{\it Symbolic Algebraic Comp.\ ISSAC~}\jref{\bf #1}{#2}{#3}, \url{https://dx.doi.org/#4}}
\def\issac#1#2#3#4{{\it Proc.\ Int.\ Symp.\ Symbolic Algebraic~Comp.~}\jref{\bf #1}{#2}{#3}, \url{https://dx.doi.org/#4}}
\def\icms#1#2#3#4{{\it Proc.\ Int.\ Congress Mathe.\ Software~}\jref{\bf #1}{#2}{#3}, \url{https://dx.doi.org/#4}}
\def\ipl#1#2#3#4{{\it Inf.~Process.~Lett.~}\jref{\bf #1}{#2}{#3}, \url{https://dx.doi.org/#4}}
\def\jsc#1#2#3#4{{\it J.~Symb.~Comp.~}\jref{\bf #1}{#2}{#3}, \url{https://dx.doi.org/#4}}
\def\macis#1#2#3#4{{\it Math.\ Aspects Comp.\ Information Sci.~}\jref{\bf #1}{#2}{#3}, \url{https://dx.doi.org/#4}}

\def\pasco#1#2#3#4{{\it Proc.\ Int.\ Workshop Parallel Symbolic Comp.~}\jref{\bf #1}{#2}{#3}, \url{https://dx.doi.org/#4}}

\def\smjcat#1#2#3#4{{\it SIAM~J.~Comp.~}\jref{\bf #1}{#2}{#3}, \url{https://dx.doi.org/#4}}
\def\stoc#1#2#3#4{{\it Proc.\ ACM Symp.\ Theory Comp.~}\jref{\bf #1}{#2}{#3}, \url{https://dx.doi.org/#4}}
\def\symsac#1#2#3#4{{\it Proc.\ ACM Symp.\ Symbolic Algebraic Comp.~}\jref{\bf #1}{#2}{#3}, \url{https://dx.doi.org/#4}}
\def\tcs#1#2#3#4{{\it Theor.~Comp.\ Sci.~}\jref{\bf #1}{#2}{#3}, \url{https://dx.doi.org/#4}}
\newcommand{\jref}[3]{{\bf #1} (#2) #3}
\newcommand{\hepph}[1]{\href{https://arXiv.org/abs/hep-ph/#1}{\texttt{hep-ph/#1}}}

\newcommand{\arxiv}[2]{\href{https://arXiv.org/abs/#1}{\texttt{arXiv:#1\,[#2]}}}
\newcommand{\bibentry}[4]{#1, {\it #2}, #3\ifthenelse{\equal{#4}{}}{}{, }#4.}

\title{Interpolation of Dense and Sparse Rational Functions and other Improvements in \code{FireFly}}
\author{Jonas Klappert, Sven Yannick Klein, and Fabian Lange}
\affil{Institute for Theoretical Particle Physics and Cosmology, RWTH
  Aachen University, D-52056 Aachen, Germany}
\date{}

\begin{document}
\maketitle
\thispagestyle{empty}
\begin{abstract}
We present the main improvements and new features in version \code{2.0} of the open-source \code{C++} library \code{FireFly} for the interpolation of rational functions.
This includes algorithmic improvements, e.g.\ a hybrid algorithm for dense and sparse rational functions and an algorithm to identify and remove univariate factors.
The new version is applied to a Feynman-integral reduction to showcase the runtime improvements achieved.
Moreover, \code{FireFly} now supports parallelization with \code{MPI} and offers new tools like a parser for expressions or an executable for the insertion of replacement tables.
\end{abstract}

\begin{textblock*}{10em}(\textwidth,1.5cm)
\raggedright\noindent
\pheaderline
\end{textblock*}

\clearpage
\pagenumbering{roman}

\section*{NEW VERSION PROGRAM SUMMARY}
\textit{Program title:} \code{FireFly}

\textit{Licensing provisions:} \abbrev{GNU} General Public License 3 (\abbrev{GPL})

\textit{Programming language:} \code{C++}

\textit{Supplementary material:} This article

\textit{Journal Reference of previous version:} J.~Klappert, F.~Lange, \textit{Reconstructing Rational Functions with \texttt{FireFly}}, \cpc{247}{2020}{106951}{10.1016/j.cpc.2019.106951}, \arxiv{1904.00009}{cs.SC}.

\textit{Does the new version supersede the previous version?:} Yes

\textit{Reasons for the new version:} Significant performance improvements and new features

\textit{Summary of revisions:} We implemented new algorithms: The racing algorithm of Ref.\,[1] for univariate polynomials, a dense and sparse hybrid algorithm for rational functions, and an algorithm to search for univariate factors which can be removed in the actual interpolation.
In addition, we changed the interface to allow for an overhead reduction inspired by vectorization and implemented the parallelization with \code{MPI}.
Moreover, we include some new tools, e.g.\ a parser for expressions and an executable for the insertion of replacement tables.

\textit{Nature of problem:} The interpolation of an unknown rational function, called black box, from only its evaluations can be used in many physical contexts where algebraic calculations fail due to memory and runtime restrictions.

\textit{Solution method:}
The black-box function is evaluated at different points over a finite field. These points are then used by interpolation algorithms\,[1-4] to obtain the analytic form of the function. The elements of a finite field are promoted to $\mathbb{Q}$ using rational reconstruction algorithms\,[5,6].

\textit{Additional comments including Restrictions and Unusual features:} For better performance, we advise to use \code{FLINT}\,[7] for the finite-field arithmetics and an improved memory allocator like \code{jemalloc}\,[8].

\textit{References:}\\
$[1]$ E.~Kaltofen, W.-s.~Lee, \textit{On Exact and Approximate Interpolation of Sparse Rational Functions}, \jsc{36}{2003}{365--400}{10.1016/S0747-7171(03)00088-9}.\\
$[2]$ M.~Ben-Or, P.~Tiwari, \textit{A Deterministic Algorithm for Sparse Multivariate Polynomial Interpolation}, \stoc{20}{1988}{301--309}{10.1145/62212.62241}.\\
$[3]$ R.~Zippel, \textit{Interpolating Polynomials from their Values}, \jsc{9}{1990}{375--403}{10.1016/S0747-7171(08)80018-1}.\\
$[4]$ A.~Cuyt, W.-s.~Lee, \textit{Sparse interpolation of multivariate rational functions}, \tcs{412}{2011}{1445--1456}{10.1016/j.tcs.2010.11.050}.\\
$[5]$ P.S.~Wang, \textit{A p-adic Algorithm for Univariate Partial Fractions}, \symsac{1981}{1981}{212--217}{10.1145/800206.806398}.\\
$[6]$ M.~Monagan, \textit{Maximal Quotient Rational Reconstruction: An Almost Optimal Algorithm for Rational Reconstruction}, \issac{2004}{2004}{243--249}{10.1145/1005285.1005321}.\\
$[7]$ W.~Hart et al., \textit{FLINT: Fast Library for Number Theory}, \href{http://www.flintlib.org/}{\texttt{http://www.flintlib.org/}}.\\
$[8]$ J.~Evans et~al., \textit{jemalloc -- memory allocator}, \href{http://jemalloc.net/}{\texttt{http://jemalloc.net/}}.\\

\clearpage
\tableofcontents
\clearpage
\pagenumbering{arabic}
\setcounter{page}{1}

\section{Introduction}
The general interpolation of unknown polynomials based on their evaluations, i.e.\ \emph{probes}, dates back to Newton's research in the 17$^\mathrm{th}$ century and the algorithm named after him\,\cite{von_zur_Gathen}.
This problem class is known as \emph{black-box} interpolation.
In the last fifty years, several sophisticated interpolation algorithms for multivariate polynomials have been developed, improved, and combined, e.g.\ in \citeres{Zippel:1979,Zippel:1990,Ben-Or:1988,Kaltofen:1988,Kaltofen:2000,Kaltofen:2003,Javadi:2010}.
Some of the algorithms are designed for \emph{dense} polynomials where most of the monomials up to the degree of the polynomial are nonzero, some are designed for \emph{sparse} polynomials where only few monomials are nonzero, and some for both.
In contrast, the interpolation of rational functions is a younger field of research dating back to Thiele's formula from the beginning of the 20$^\mathrm{th}$ century\,\cite{von_zur_Gathen}.
The first algorithms for multivariate rational functions were developed thirty years ago\,\cite{Kaltofen:1990_1,Grigoriev:1990,Grigoriev:1991,Grigoriev:1994} and since then more algorithms have been published\,\cite{deKleine:2005,Kaltofen:2007,Cuyt:2011,Huang:2017}.
Most of them rely on one or more polynomial-interpolation algorithms as sub-algorithms.
Again, some of the algorithms are designed for sparse and others for dense functions.
Moreover, most of the algorithms mentioned above are designed with finite-field arithmetic in mind, e.g.\ integers modulo a prime number.
The elements of a finite field can be promoted to rational numbers with rational reconstruction algorithms\,\cite{Wang:1981,Monagan:2004}.

Last year, two of the authors of the present paper published the \code{C++} library \code{FireFly} for the interpolation of rational functions\,\cite{Klappert:2019emp} based on the sparse rational-function-interpolation algorithm by Cuyt and Lee\,\cite{Cuyt:2011} with an embedded Zippel algorithm for the polynomials\,\cite{Zippel:1979,Zippel:1990}.
Shortly after, the \code{C++} library \code{FiniteFlow} was published\,\cite{Peraro:2019svx}.
Even though it uses the central idea of \citere{Cuyt:2011}, it is a dense implementation.

In this paper, we present version \code{2.0} of \code{FireFly} which includes both algorithmic and technical improvements.
The algorithmic improvements are presented in \sct{sec:algorithmic_improvements}.
First, we embedded the univariate racing algorithm of Kaltofen, Lee, and Lobo\,\cite{Kaltofen:2000,Kaltofen:2003} into the multivariate Zippel algorithm.
It races the dense Newton algorithm against the sparse Ben-Or/Tiwari algorithm\,\cite{Ben-Or:1988}.
This significantly improves the performance of \code{FireFly} for sparse functions.
Secondly, we present a modified version of the algorithm of \citere{Cuyt:2011} based on the \emph{temporary pruning} idea of \citeres{Kaltofen:2000,Kaltofen:2003}.
Instead of only pruning on the level of polynomials, we now also prune on the level of rational functions.
This increases the performance especially for dense functions.
Lastly, we implemented an algorithm to search for univariate factors of the rational function before the actual multivariate interpolation.
By removing those factors, the interpolation can become much simpler.
In the context of integration-by-parts (\abbrev{IBP}) reductions in theoretical particle physics\,\cite{Tkachov:1981wb,Chetyrkin:1981qh}, similar ideas have been pursued in \citeres{Smirnov:2020quc,Usovitsch:2020jrk}.

In \sct{sec:Technical_Improvements} we summarize the main technical improvements since the release of \code{FireFly}.
First, we changed the interface to pave the way for a reduction of overheads inspired by the vectorization of modern \abbrev{CPU}s.
Instead of evaluating the black box at one parameter point, one can evaluate it at several points at the same time.
This reduces the overhead of many problems, most notably when solving systems of equations.
Secondly, we implemented a parser which allows the user to read in expressions
from files and strings and use them as building blocks for the black box.
The parser is designed for fast evaluations.
Lastly, \code{FireFly} now supports the parallelization with \code{MPI}.
This enables \code{FireFly} to utilize computer clusters for the black-box evaluations.
Since they are the main bottleneck of interpolation problems in a physical context, this can significantly speed up calculations.

In precision high-energy physics, integral reductions based on \abbrev{IBP} relations\,\cite{Tkachov:1981wb,Chetyrkin:1981qh} play a major role.
The most prominent strategy is the Laporta algorithm\,\cite{Laporta:2001dd}, which is implemented in several public\,\cite{Anastasiou:2004vj,Smirnov:2008iw,Smirnov:2013dia,Smirnov:2014hma,Smirnov:2019qkx,Studerus:2009ye,vonManteuffel:2012np,Maierhoefer:2017hyi,Maierhofer:2018gpa} and numerous private codes.
In the last decade, finite-field and interpolation techniques spread to this field beginning with \citere{Kauers:2008zz}.
The former can be used to remove redundant information to reduce the size of the system of equations\,\cite{Kant:2013vta} as implemented in \code{Kira}\,\cite{Maierhoefer:2017hyi,Maierhofer:2018gpa}.
By directly interpolating the final result, one can also circumvent the large intermediate expressions which usually appear\,\cite{vonManteuffel:2014ixa,Peraro:2016wsq}.
\code{FIRE6} is the first public tool available with interpolation techniques implemented for \abbrev{IBP} reductions\,\cite{Smirnov:2019qkx}.
These techniques also facilitate the development of new strategies for \abbrev{IBP} reductions, see e.g.\ \citeres{Bendle:2019csk,Guan:2019bcx}.
Another bottleneck can be the insertion of the final reduction tables into expressions like amplitudes.
Also, this problem might be eased with the help of finite-field arithmetic and interpolation techniques.

In \sct{sec:IBPs} we apply the current development version of \code{Kira} to a reduction problem to illustrate the improved performance of \code{FireFly\;2.0} and describe a tool based on \code{FireFly\;2.0}, which can insert replacement tables into expressions.

Recently, a similar strategy to obtain amplitudes from floating-point probes has been presented in \citeres{DeLaurentis:2019phz,DeLaurentis:2019vkf}.

\section{Algorithmic improvements}
\label{sec:algorithmic_improvements}
In the following sections we describe the algorithmic improvements that have
been implemented since the release of the first version of \code{FireFly}. We
refer the reader to \citere{Klappert:2019emp} and the references therein for a
description of the base algorithms implemented in \code{FireFly} and our notation.
The interpolation algorithms presented here are discussed over a prime field $\zp{p}$ with characteristic $p$, where $p$ is its defining prime.

\subsection{Racing Newton vs Ben-Or/Tiwari}
\label{sec:Newton_vs_Ben-Or/Tiwari}
As sub-algorithm for the interpolation of rational functions with \code{FireFly} we employ a multivariate polynomial interpolation with the Zippel algorithm \cite{Zippel:1979,Zippel:1990}.
The Zippel algorithm itself requires a univariate polynomial-interpolation algorithm.
Instead of a univariate interpolation with the Newton algorithm, we now employ the racing algorithm presented in \citeres{Kaltofen:2000,Kaltofen:2003}.
It races the dense Newton interpolation against the Ben-Or/Tiwari algorithm\,\cite{Ben-Or:1988}.
The Ben-Or/Tiwari algorithm is a sparse polynomial-interpolation algorithm and scales with the number of terms, whereas the Newton interpolation scales with the degree of the polynomial.
The racing algorithm updates both interpolation algorithms sequentially with a new probe until either of them terminates.
Hence, it performs well for both sparse and dense polynomials.

In the following sections, we summarize both the univariate Ben-Or/Tiwari algorithm and its sub-algorithms based on \citeres{Kaltofen:1988,Kaltofen:2000} and the racing algorithm of \citeres{Kaltofen:2000,Kaltofen:2003}.

\subsubsection{Ben-Or/Tiwari algorithm}
\label{sec:Ben-Or/Tiwari}
The Ben-Or/Tiwari algorithm is an interpolation algorithm for multivariate Polynomials\,\cite{Ben-Or:1988}.
However, for our purpose the univariate case is sufficient.
Consider the polynomial
\begin{equation}
  \label{eq:univariate_poly_sparse}
  f(z) = \sum_{i=1}^{T}c_{\alpha_i} z^{\alpha_i}
\end{equation}
with $T$ nonzero terms. The $c_{\alpha_i}$ are the coefficients of the monomials of degrees $\alpha_i$.
For an arbitrary element $y\in\zp{p}$, the anchor point, we then define the evaluations of the polynomial at integer powers $i\geq1$ of $y$ as
\begin{equation}
  a_i = f\big(y^i\big) \,.
\end{equation}
The original formulation of the algorithm uses $i\geq0$.
However, then it is more vulnerable to accidental cancellations, because the first evaluation is performed at $y^0 = 1 \ \forall y$\,\cite{Hu:2016}.

The main ingredient of the Ben-Or/Tiwari algorithm is the auxiliary polynomial
\begin{equation}
  \label{eq:aux_poly}
  \zeta(z) = \prod_{i=1}^T\left(z-y^{\alpha_i}\right) = z^T + \lambda_{T-1} z^{T-1} + \ldots + \lambda_0 \,,
\end{equation}
with the coefficients $\lambda_k$.
It is designed such that one can obtain the degrees $\alpha_i$ of the polynomial $f(z)$ by computing the roots of $\zeta(z)$. Hence, by definition,
\begin{equation}
  \sum_{i=0}^T c_{\alpha_i} \left(y^ {\alpha_i} \right)^j \zeta\big( y^{\alpha_i} \big) = 0
\end{equation}
holds for any integer $j\geq0$.
On the other hand, this sum can be rewritten as
\begin{equation}
  \begin{split}
    \sum_{i=0}^T c_{\alpha_i} \left(y^ {\alpha_i} \right)^j \zeta\big( y^{\alpha_i} \big) &= \sum_{k=0}^{T-1} \lambda_k \left( \sum_{i=0}^T c_{\alpha_i} \left( y^{\alpha_i} \right)^{j+k} \right) + \sum_{i=0}^T c_{\alpha_i} \left( y^{\alpha_i} \right)^{j+T} \\
    &= \sum_{k=0}^{T-1} \lambda_k f \big( y^{j+k} \big) + f \big( y^{j+T} \big) \,,
  \end{split}
\end{equation}
with the definition of $f(z)$ in \eqn{eq:univariate_poly_sparse}.
Therefore, one obtains the linear relation
\begin{equation}
  0 = \sum_{k=0}^{T-1} \lambda_k a_{j+k} + a_{j+T}
  \label{eq:lfsr}
\end{equation}
between the coefficients $\lambda_k$ of the auxiliary polynomial.
This allows for the construction of a system of equations based on the evaluations $a_{j+k}$ of the polynomial $f(z)$.
Since there are exactly $T$ unknowns, $T$ linearly independent equations of this form are required.
The minimum number of evaluations can be achieved by constructing one equation with $T+1$ evaluations $a_{j+k}$ and reusing $T$ of them for the next equation.
Thus, one only has to compute one additional probe for each equation after the first.
Therefore, in total $2T$ probes are required to find the auxiliary polynomial $\zeta(z)$.
The system of $T$ equations can then be written in matrix form as
\begin{equation}
  \label{eq:Hankel_system}
  \begin{pmatrix}
    a_1 & a_2 & \cdots & a_{T} \\
    a_2 & a_3 & \cdots & a_{T+1} \\
    \vdots & \vdots & \ddots & \vdots \\
    a_{T} & a_{T+1} & \cdots & a_{2T-1}
  \end{pmatrix}
  \begin{pmatrix}
    \lambda_0 \\
    \lambda_1 \\
    \vdots \\
    \lambda_{T-1}
  \end{pmatrix}
  = -
  \begin{pmatrix}
    a_{T+1} \\
    a_{T+2} \\
    \vdots \\
    a_{2T}
  \end{pmatrix}
  \,.
\end{equation}
\eqn{eq:Hankel_system} is a Hankel system which can be solved iteratively with the Berlekamp-Massey algorithm (Alg.\,(\ref{alg:Berlekamp-Massey})) described in the next section.
This can be understood by the fact that the $\lambda_k$ of \eqn{eq:lfsr} build a linear-feedback shift register that generates the sequence $a_i$.

Once the auxiliary polynomial of \eqn{eq:aux_poly} has been calculated, its roots have to be obtained.
Since the roots are integer powers $i$ of $y$ by construction, the simplest
method is to evaluate $\zeta(z)$ at these values and check whether the result
is zero.
Hence, one starts with $y^{i=0}=1$ and increases the power $i$ until the number of roots matches the degree of $\zeta(z)$.
The powers $i$ corresponding to the roots are the degrees $\alpha_i$ of the monomials of $f(z)$ and the number of roots corresponds to the number of terms $T$.

The coefficients $c_{\alpha_i}$ of $f(z)$ can be obtained by constructing and solving a system of the equations
\begin{equation}
  \label{eq:vandermonde_system}
  a_i = f\big(y^i\big) = \sum_{j=1}^T c_{\alpha_i} \left(y^i\right)^{\alpha_i} \,.
\end{equation}
By choosing a sequence of $i$, one can construct a shifted transposed Vandermonde system for the coefficients $c_{\alpha_i}$\,\cite{Hu:2016}.
We choose the probes $1 \leq i \leq T$ in our implementation.
A solving algorithm for these systems has already been implemented in the first version of \code{FireFly}\,\cite{Klappert:2019emp}.
No additional probes are required for this step because all previously obtained probes can be reused.

The case $f(z) = 0$ is special because the Hankel matrix in
\eqn{eq:Hankel_system} becomes a null matrix.
Therefore, $\zeta(z)$ is undetermined.
However, the Berlekamp-Massey algorithm yields $\zeta(z)=1$, which has no roots.
Since this is the only case without roots, one can easily intercept it.

Since the interpolation is performed over finite fields, one has to account for the fact that the element $y$ may generate a cyclic group with a small order, i.e.\ there is a small power $b$ for which $y^b = 1$.
Hence, if $y^j$ is a root of the auxiliary polynomial also $y^{\left(j+n\cdot b\right)}\, \forall n \in \mathbb{N}$ are roots of the auxiliary polynomial, which is problematic if $b$ is smaller than the degree of the polynomial.
Additionally, if the power $b$ is smaller than the number of terms $T$, the equations in the Hankel system of \eqn{eq:Hankel_system} and the Vandermonde system are not linearly independent. Therefore, it is necessary to use fields with a large characteristic
so that the integer $b$ is larger than the degree and the number of terms of the interpolated
polynomial for the anchor point $y$ with high probability\,\cite{Kaltofen:2003}.

The Ben-Or/Tiwari algorithm can be illustrated as follows. Assume the polynomial
\begin{equation}
  f(z) = z^2 + 1
\end{equation}
defined over the field $\zp{509}$ shall be interpolated.
We choose $y=2$ as the anchor point.
Since there are $T=2$ terms, $2T=4$ probes are required:
\begin{equation}
  f\big(y^1\big) = 5 \,,\quad f\big(y^2\big) = 17 \,, \quad f\big(y^3\big) = 65 \,, \quad f\big(y^4\big) = 257 \,.
\end{equation}
With this input, the Berlekamp-Massey algorithm leads us to
\begin{equation}
  \zeta(z) = z^2+504z+4
\end{equation}
as the auxiliary polynomial.
Evaluating $\zeta(z)$ at integer powers of $y$ until two roots are found yields
\begin{equation}
  \zeta\big(y^0\big) = 0 \,, \quad \zeta\big(y^1\big) = 507 \,, \quad \zeta\big(y^2\big) = 0 \,.
\end{equation}
Thus, the polynomial consists of monomials with degrees $0$ and $2$.
Their coefficients are then obtained by solving the system
\begin{equation}
  f\big(2^1\big) = 5 = c_0 + c_2 \left(2^1\right)^2 \,, \quad f\big(2^2\big) = 17 = c_0 + c_2 \left(2^2\right)^2 \,,
\end{equation}
which yields $c_0 = c_2 = 1$.

\subsubsection{Berlekamp-Massey algorithm}
The Berlekamp-Massey algorithm was originally a decoding algorithm in coding theory designed by Berlekamp\,\cite{Berlekamp}.
Massey simplified and connected it to linear-feedback shift registers\,\cite{Massey:1969}.
Later, it was interpreted in terms of Hankel systems\,\cite{Imamura:1987,Jonckheere:1989}.

The Berlekamp-Massey algorithm is an iterative algorithm to find the linear generator
\begin{equation}
  \Lambda(z) = \lambda_0 z^T +\ldots + \lambda_{T-1} z + \lambda_T
  \label{eq:lin_gen}
\end{equation}
of a stream of elements $a_1,a_2,\ldots$ given by \eqn{eq:lfsr} over an arbitrary field. Although the sequence is unbounded, the algorithm will compute the polynomial $\Lambda(z)$ after processing $2T$ elements from the input stream.
The degree of $\Lambda(z)$ is the length $L$ of the linear-feedback shift register. While processing a new $a_r$ as a new iteration step, \eqn{eq:lfsr} is checked, i.e.\ the r.h.s.\ of \eqn{eq:lfsr} is evaluated using the polynomial $\Lambda_{r-1}(z)$ obtained in the previous iteration step of the algorithm. This yields the discrepancy
\begin{equation}
    \Delta_r \equiv \sum_{k=0}^{L-1} \lambda_k a_{j+k} + a_{j+L}\,,
\end{equation}
with the $\lambda_k$ being the coefficients of the current polynomial $\Lambda_{r-1}(z)$ and $j = r-L$. Note that $\lambda_L = 1$ always holds.
If $\Delta_r$ vanishes, the polynomial $\Lambda_{r-1}(z)$ is a valid generator for the sequence of $a_i$ up to $a_r$ and does not need to be changed. If $\Delta_r \neq 0$, there are two update possibilities depending of the current length $L_{r-1}$. The first update increases the generator's degree and the second update adjusts the lower orders of the generator's coefficients.
The details of the update steps are summarized in \alg{alg:Berlekamp-Massey}.

\begin{algorithm}
\caption{Berlekamp-Massey algorithm from \citere{Kaltofen:2000}.}
\label{alg:Berlekamp-Massey}
\algrenewcommand\algorithmicrequire{\textbf{Input:}}
\algrenewcommand\algorithmicensure{\textbf{Output:}}
\begin{algorithmic}
\Require $a_1,a_2,\ldots$
\State Initialization: $\Lambda_0(z)\leftarrow 1$; $B_0(z)\leftarrow 0$; $L_0\leftarrow0$; $\Delta\leftarrow1$;
\For{$r=1,  2, \dots $}
\State (Calculate the discrepancy $\Delta_r$, assuming that $\Lambda_{r-1}(z) = \lambda_s + \lambda_{s-1} z + \dots + \lambda_0 z^s$)
\State $\Delta_r \leftarrow \lambda_s a_{r} + \lambda_{s-1} a_{r-1} + \dots + \lambda_0 a_{r-s}$;
\If{$\Delta_r=0$}
\State $\Lambda_r(z) \gets \Lambda_{r-1}(z)$; $B_r(z)\gets z B_{r-1}(z)$; $L_r \gets L_{r-1}$;
\ElsIf{$\Delta_r \neq 0$ and $2L_{r-1} < r$}
\State $B_{r}(z)\gets \Lambda_{r-1}(z)$; $\Lambda_r(z)\gets \Lambda_{r-1}(z)-\tfrac{\Delta_r}{\Delta} z B_{r-1}(z)$; $L_r\gets r-L_{r-1}$; $\Delta\gets\Delta_r$;
\ElsIf{$\Delta_r \neq 0$ and $2L_{r-1} \geq r$}
\State $\Lambda_r(z)\gets \Lambda_{r-1}-\tfrac{\Delta_r}{\Delta}zB_{r-1}(z)$; $B_{r}\gets zB_{r-1}(z)$; $L_r\gets L_{r-1}$;
\EndIf
\EndFor
\end{algorithmic}
\end{algorithm}

Thus, the Berlekamp-Massey algorithm determines the coefficients $\lambda_k$ in \eqn{eq:lfsr}, which is equivalent to solving the Hankel system given by \eqn{eq:Hankel_system}\,\cite{Imamura:1987,Jonckheere:1989}.
The auxiliary polynomial $\zeta(z)$ of the Ben-Or/Tiwari algorithm can be obtained by reversing the coefficients in $\Lambda(z)$, i.e.\
\begin{equation}
  \zeta(z) = \lambda_T z^T +\ldots + \lambda_{1} z + \lambda_0
\end{equation}
with $\lambda_T\equiv 1$.

\subsubsection{Early termination in the Ben-Or/Tiwari algorithm}
\label{sec:early_termination}
The Ben-Or/Tiwari algorithm described in \sct{sec:Ben-Or/Tiwari} requires the number of terms $T$ as input parameter.
However, in \citeres{Kaltofen:2000,Kaltofen:2003} it was formulated in an \emph{early termination} version, which does not rely on $T$ as input.

The first step of the Ben-Or/Tiwari interpolation is the Berlekamp-Massey algorithm to compute the auxiliary polynomial $\zeta(z)$.
\alg{alg:Berlekamp-Massey} processes one probe after another and does not rely on subsequent probes.
If the discrepancy $\Delta_r = 0$, $\Lambda_{r-1}(z)$ already solves the Hankel system of \eqn{eq:Hankel_system}.
This happens either accidentally or if the number of probes reaches $2T+1$.
The probability of the former is less than\,\cite{Kaltofen:2000,Kaltofen:2003}
\begin{equation}
  \dfrac{T\left(T+1\right)\left(2T+1\right)D}{6\cdot \#\mathbb{F}} \,,
\end{equation}
where $D$ is the degree of the polynomial and $\#\mathbb{F}$ is the characteristic of the field $\mathbb{F}$.

Thus, the termination criterion is to terminate after a sequence of $\eta$ probes yields $\Delta_r = 0$ and the total number of probes with non-vanishing discrepancy corresponds to a Hankel system, i.e.\ $r = 2L+\eta$ with $L\in\mathbb{N}$.
In our experience, it is sufficient to choose $\eta = 1$, when interpolating over fields with 63-bit characteristics.
With the auxiliary polynomial $\zeta(z)$ known, the Ben-Or/Tiwari algorithm can proceed.
By definition, $\zeta(z)$ has to completely factorize into mutually distinct factors $z - y^i$.
There is an upper bound on the power $i$ if there is an upper bound on the degree of the polynomial, which is the case when embedding the Ben-Or/Tiwari interpolation into the interpolation of a rational function following the strategy of \citere{Cuyt:2011} (see \sct{sec:hybrid_racer} for details).
If the auxiliary polynomial completely factorizes in this way, the early termination was correct with high probability.
If the factorization fails, we take the result of the Newton interpolation
which always terminates.

\subsubsection{Racing}
As proposed in \citere{Kaltofen:2000,Kaltofen:2003}, we race the Newton and the Ben-Or/Tiwari algorithm against each other.
Our implementation of the former and its early termination criterion are described in \citere{Klappert:2019emp}.
For each probe, both algorithms get updated with this numerical evaluation.
If either of them reaches its termination criterion, we abort the other interpolation and take the result of the terminated algorithm.

The advantage of the racing algorithm is that it minimizes the number of probes required to interpolate a polynomial.
The Newton interpolation is a dense algorithm and takes $D+1+\eta$ probes with the termination criterion, i.e.\ one probe for the term of each degree up to the maximum degree $D$ plus $\eta$ additional checks.
On the other hand, the Ben-Or/Tiwari interpolation finishes after $2T+\eta$ probes as described in the previous section and terminates much earlier than Newton's interpolation if $T\ll D$.
Therefore, the racing algorithm scales well for both dense and sparse polynomials.
Moreover, when embedding the racing algorithm into the multivariate Zippel algorithm, it usually scales better than both algorithms alone, because a different algorithm can win for different variables\,\cite{Kaltofen:2000,Kaltofen:2003}.

This improvement comes with additional runtime for the Ben-Or/Tiwari algorithm as the only drawback.
However, this additional runtime is regained manifold if the Ben-Or/Tiwari
interpolation finishes faster than Newton's interpolation sometimes and the
runtime for each probe dominates the total runtime of the interpolation.

\subsubsection{Comparison of Newton and racing algorithm}
To illustrate the changes in the number of required black-box probes, we compare \code{FireFly} \code{1.1} 
with and without the racing algorithm. When no racing algorithm is used, the Newton interpolation is used by default. We test the benchmark functions
\begin{align}
f_1(z_1,\dots,z_{20}) &= \frac{\sum_{i=1}^{20}z_i^{20}}{\sum_{i=1}^{10}z_i^{20}-\sum_{i=11}^{20}z_i^{20}}\;,
\label{eq:hr_b_1}
\\
f_2(z_1,\dots,z_5) &= \frac{z_1^{100}+z_2^{200}+z_3^{300}}{z_1z_2z_3z_4z_5+z_1^4z_2^4z_3^4z_4^4z_5^4}\;,
\label{eq:hr_b_2}
\\
f_3(z_1,\dots,z_5) &= \frac{\left(1+z_1+z_2+z_3+z_4+z_5\right)^{17}-1}{z_4-z_2+z_1^{10}z_2^{10}z_3^{10}z_4^{10}z_5^{10}}\;
\label{eq:hr_b_3}
\end{align}
and enable the scan for a sparse shift by default. The results are summarized in \tab{tab:nbt_bench}.
\begin{table}[ht]
\begin{center}
\caption{Comparison of Newton and racing in \code{FireFly\;1.1} with
  respect to runtime and required probes for the benchmark functions defined
  by \eqs{eq:hr_b_1}--(\ref{eq:hr_b_3}). The interpolation is performed over $\zp{9223372036854775783}$ using a single thread of an \abbrev{AMD} Ryzen 5 3400G processor.
  The shift scan has been used before to find a sparse shift for each function.}
\label{tab:nbt_bench}
\begin{tabular}{c|c c|c c}
\toprule
\multicolumn{1}{c|}{} & \multicolumn{2}{c|}{Newton} & \multicolumn{2}{c}{Racing}\\
Function & Probes & Runtime & Probes & Runtime\\
\midrule
$f_1$ & 15242 & 0.33\,s & 3082 & 0.15\,s \\
$f_2$ & 134906 & 30\,s & 1957 & 0.6\,s  \\
$f_3$ & 102358 & 20\,s & 101926 & 20\,s \\
\end{tabular}
\end{center}
\end{table}

Since the Newton interpolation is a dense interpolation, it requires up to two
orders of magnitude more probes compared to the racing algorithm for sparse
functions like $f_1$ and $f_2$. Consequently, the racing algorithm can improve
the runtime significantly in such cases. For dense functions like $f_3$, there
is almost no difference between the two algorithms in the number of required
black-box probes.
Therefore, the additional runtime for the Ben-Or/Tiwari algorithm is negligible in most applications, especially if the computation of the probes is dominant compared to the runtime of the interpolation.

Since \code{FireFly\;2.0.3}, files with the functions (\ref{eq:hr_b_1})--(\ref{eq:hr_b_3}) as well as executables are shipped with the source code.
We refer to the file \code{README.md} accompanying the source code for instructions on how to compile and run them.

\subsection{Interpolation of sparse and dense multivariate rational functions}
\label{sec:hybrid_racer}
In this section, we present a modified version of the interpolation algorithm developed by Cuyt and Lee\,\cite{Cuyt:2011} that improves the scaling behavior for dense rational functions.
Our new algorithm is designed to find a balance between the required number of black-box probes for dense and sparse rational functions.

In practice, it is based on the \emph{temporary pruning} idea of \citeres{Kaltofen:2000,Kaltofen:2003}, i.e.\ the coefficients which have already been interpolated are removed from the system of equations.
Consequently, we race a dense interpolation, i.e.\ including the effects of the variable shift, against the sparse approach by \citere{Cuyt:2011}.
The algorithm can be divided into the following steps:
\begin{enumerate}
\item Find a variable shift $\vec{s}=(s_1,\dots,s_n)$ for a given black-box function $f(z_1,\ldots,z_n)$ of $n$ variables, that generates a constant term in either numerator or denominator and involves a minimal set of nonzero $s_i,i\in[1,n]$.
\item Interpolate the univariate auxiliary function
\begin{equation}
\tilde{f}(t\vec{z}+\vec{s})= f(tz_1 + s_1,\dots,tz_n+s_n)
\end{equation}
in the homogenization variable $t$\,\cite{Diaz:1998} at numerical values for $\vec{z}$.
One can set $z_1=1$ and restore its dependence later by homogenization.
The interpolation of $\tilde{f}$ is only performed once and then replaced by solving a system of equations containing all nonzero degrees in $t$ that still need to be processed by multivariate polynomial interpolations.
\item\label{it:poly_int} Process all coefficients of $\tilde{f}$ by multivariate polynomial interpolations simultaneously including the effects of the shift.
Hence, all but the highest degree terms in numerator and denominator are interpolated densely.
\item When an interpolation terminates, restore its dependence in $z_1$ by homogenization.
When the interpolation of the polynomial coefficient with the currently highest degree of $\tilde{f}$, labeled $f_h$, in either numerator or denominator terminates, abort the interpolation of the polynomial coefficient of the next-to-highest degree, labeled $f_l$.
Calculate the monomials that occur by shifting $f_h$ and restart the interpolation of $f_l$ while subtracting the shift monomials of all higher degree terms numerically.
Hence, this is now a sparse interpolation.
The probes processed during the dense interpolation can be reused.
If the stored probes are not sufficient for the restarted interpolation to terminate, proceed.
\item Prune the system of equations for the auxiliary function $\tilde{f}$ by all interpolated polynomial coefficients.
When a coefficient has been interpolated sparsely, i.e.\ the monomials created by shifts in higher degree polynomials were subtracted, re-add those monomials to the system of equations for consistency.
While pruning the system of equations, the already solved polynomials created by the shift have to be evaluated at $t\vec{z}$ instead of $t\vec{z}+\vec{s}$.
\item After all interpolations terminated, subtract all polynomials generated by shifts of higher degree coefficients from the coefficients which have been interpolated densely.
\end{enumerate}
This algorithm is applicable for both dense and sparse rational functions and avoids oversampling for dense functions, which is apparent in the algorithm of \citere{Cuyt:2011}, while still maintaining a good performance for sparse functions. For further reference, we call this algorithm \emph{hybrid racer}, because it races sparse and dense interpolations while being efficient for both cases.

In our implementation, we use the multivariate Zippel algorithm as described in \citere{Klappert:2019emp} with the embedded univariate racing algorithm described in \sct{sec:Newton_vs_Ben-Or/Tiwari} for the polynomial interpolation in point\,\ref{it:poly_int}.
Since we interpolate the functional dependence of each variable sequentially ($z_2\to z_3\to \dots \to z_n$) and the polynomial coefficients proceed differently in general, we have to choose for which coefficient we compute the next probes.
We prefer to schedule probes for $z_i$, where $i$ is the index closest to $n$.
This ensures that the total number of probes is bounded by the maximum number of possible terms of a dense rational function with given degrees.

To illustrate this algorithm, consider the black-box function
\begin{equation}
f(z_1,z_2) = \frac{z_1z_2^3+z_1^2z_2^2+z_1^3z_2+z_1^4+z_2^5}{z_2} \,.
\label{eq:hr_f}
\end{equation}
To fix the normalization, we shift $z_2$ as $z_2+2$ yielding
\begin{equation}
f(z_1,z_2+2) = \frac{z_1(z_2+2)^3+z_1^2(z_2+2)^2+z_1^3(z_2+2)+z_1^4+(z_2+2)^5}{2+z_2}\,.
\label{eq:hr_fs}
\end{equation}
After performing one interpolation of \eqn{eq:hr_fs} as a function of $t$ with eleven probes, one obtains the skeleton of the auxiliary rational function
\begin{equation}
\tilde{f}(t, tz_2 + 2) = \frac{n_{\alpha_{0}} + n_{\alpha_{1}} t + n_{\alpha_{2}} t^2 + n_{\alpha_{3}} t^3 + n_{\alpha_{4}} t^4 + n_{\alpha_{5}} t^5}{1 + d_{\beta_{1}}t}\,,
\end{equation}
where each $n_{\alpha}$ and $d_\beta$ is a multivariate polynomial, i.e.\ $n_{\alpha} \equiv n_{\alpha}(z_1, z_2)$.
Hence, we can build a system of equations, where each equation is structured as
\begin{equation}
n_{\alpha_{1,\mathrm{u}}} t + n_{\alpha_{2,\mathrm{u}}} t^2 + n_{\alpha_{3,\mathrm{u}}} t^3 + n_{\alpha_{4,\mathrm{u}}} t^4 + n_{\alpha_{5,\mathrm{u}}} t^5 - d_{\beta_{1,\mathrm{u}}}t f(t,tz_2+2) = -n_{\alpha_{0,\mathrm{s}}} + f(t,tz_2+2)\,.
\label{eq:hr_seq1}
\end{equation}
The index s (u) indicates a solved (unsolved) coefficient.
After the interpolation in $t$, the constant in the numerator, $n_{\alpha_{0},\mathrm{s}}$, is already determined and can be used to prune the system by one term.
Thus we need only six black-box probes to solve the system of equations defined by \eqn{eq:hr_seq1}, whereas the approach by \citere{Cuyt:2011} requires seven.
After solving one system built by \eqn{eq:hr_seq1}, the linear terms
$n_{\alpha_{1}}$ and $d_{\alpha_{1}}$ terminate their interpolations by reaching the degree bound in Newton's algorithm.
Therefore, the system can be pruned to size four:
\begin{equation}
n_{\alpha_{2,\mathrm{u}}} t^2 + n_{\alpha_{3,\mathrm{u}}} t^3 + n_{\alpha_{4,\mathrm{u}}} t^4 + n_{\alpha_{5,\mathrm{u}}} t^5  = -n_{\alpha_{0,\mathrm{s}}} - n_{\alpha_{1,\mathrm{s}}} t  + (1 + d_{\beta_{1,\mathrm{s}}}t )f(t,tz_2+2)\,.
\label{eq:hr_seq2}
\end{equation}
After probing the black box four additional times, the system built by \eqn{eq:hr_seq2} can be solved.
With these new probes, the Ben-Or/Tiwari algorithm for $n_{\alpha_{5}}$ terminates and the Newton algorithm for $n_{\alpha_{2}}$ reaches its degree bound.
Thus, the system can be pruned again and reduces to
\begin{equation}
 n_{\alpha_{3,\mathrm{u}}} t^3 + n_{\alpha_{4,\mathrm{u}}} t^4  = -n_{\alpha_{0,\mathrm{s}}} - n_{\alpha_{1,\mathrm{s}}} t -n_{\alpha_{2,\mathrm{s}}} t^2 - n_{\alpha_{5,\mathrm{s}}} t^5  + (1 + d_{\beta_{1,\mathrm{s}}}t )f(t,tz_2+2)
\label{eq:hr_seq3}
\end{equation}
with two unsolved coefficients.
Since the highest degree polynomial of the numerator,
$n_{\alpha_{5},\mathrm{s}}$, is now known, we can abort the interpolation of $n_{\alpha_{4}}$ and restart it with the previously computed values, while numerically subtracting the effects of shifting $z_2$ in $n_{\alpha_{5},\mathrm{s}}$.
After the next solution of \eqn{eq:hr_seq3}, $n_{\alpha_3}$ terminates by reaching the degree bound.
Thus, the system can be pruned to
\begin{equation}
n_{\alpha_{4,\mathrm{u}}} t^4  = -n_{\alpha_{0,\mathrm{s}}} - n_{\alpha_{1,\mathrm{s}}} t - n_{\alpha_{2,\mathrm{s}}} t^2 -  n_{\alpha_{3,\mathrm{s}}} t^3 - n_{\alpha_{5,\mathrm{s}}} t^5  + (1 + d_{\beta_{1,\mathrm{s}}}t )f(t,tz_2+2) \,.
\label{eq:hr_seq4}
\end{equation}
Solving \eqn{eq:hr_seq4}, allows the Newton interpolation of $n_{\alpha_4}$ to finish with the early termination criterion.
Afterwards, one has to calculate the shift expansion to remove its effects on all coefficients except $n_{\alpha_4}$, $n_{\alpha_5}$, and $d_{\beta_1}$.

In total, the hybrid racer algorithm requires 24 probes to interpolate the function given by \eqn{eq:hr_f}. Note that shifting both $z_1$ and $z_2$ does not increase the number of required probes. In contrast, our implementation of the algorithm presented in \citere{Cuyt:2011} in \code{FireFly\;1.1} requires 39 black-box probes in total using the same variable shift.
There, the interpolation of $\tilde{f}$ with Thiele's formula is replaced by a system of equations with seven unknowns after the first interpolation.
Depending on the sparse shift, the hybrid racer algorithm could potentially prefer to interpolate a rational function densely although the sparse approach by \citere{Cuyt:2011} would lead to fewer black-box probes. However, the imposed oversampling should still reflect the sparsity of the rational function.

To illustrate the performance of the hybrid-racer algorithm implemented in \code{FireFly\;2.0} compared to \code{FireFly\;1.1}, \tab{tab:hr_bench} shows a comparison in runtime and required black-box probes for the benchmark functions defined in \eqs{eq:hr_b_1}--(\ref{eq:hr_b_3}).
In addition, we include the function
\begin{align}
f_4(z_1,\dots,z_5) &= \frac{\left(1+z_1+z_2+z_3+z_4+z_5\right)^{20}-1}{z_4-z_2+z_1^{10}z_2^{10}z_3^{10}z_4^{10}z_5^{10}}\;.
\label{eq:hr_b_4}
\end{align}
to exemplify the scaling w.r.t.\ \eqn{eq:hr_b_3} by doubling the number of monomials.
Apart from runtime improvements during the development of \code{FireFly\;2.0}, the reduced amount of probes for dense functions is evident. Since the hybrid-racer algorithm is designed to be bounded at the maximum number of possible monomials, for certain types of sparse functions, the number of required black-box probes can increase in comparison to the algorithm of \citere{Cuyt:2011}. However, some of these drawbacks can be circumvented when combining it with the scan for factors described in \sct{sec:factors}.
\begin{table}[ht]
\begin{center}
\caption{Comparison of runtime and number of required probes for the benchmark functions defined by \eqs{eq:hr_b_1}--(\ref{eq:hr_b_3}) and \eqn{eq:hr_b_4} between \code{FireFly\;1.1} and \code{FireFly\;2.0}. The interpolation is performed over $\zp{9223372036854775783}$ using a single thread of an \abbrev{AMD} Ryzen 5 3400G processor.
The shift scan has been used before to find a sparse shift for each function.}
\label{tab:hr_bench}
\begin{tabular}{c|c c|c c}
\toprule
\multicolumn{1}{c|}{} & \multicolumn{2}{c|}{\code{FireFly\;1.1}} & \multicolumn{2}{c}{\code{FireFly\;2.0}}\\
Function & Probes & Runtime & Probes & Runtime\\
\midrule
$f_1$ & 3082 & 0.2\,s & 951 & 0.02\,s\\
$f_2$ & 1957 & 0.5\,s & 1285 & 0.3\,s\\
$f_3$ & 101926 & 20\,s & 26435 & 7.8\,s\\
$f_4$ & 212698 & 79\,s & 53231 & 28\,s\\
\end{tabular}
\end{center}
\end{table}

In addition to the functions (\ref{eq:hr_b_1})--(\ref{eq:hr_b_3}), also function (\ref{eq:hr_b_4}) is available in the source code of \code{FireFly} since version \code{2.0.3}.
We again refer to the file \code{README.md} accompanying the source code for instructions.

\subsection{Finding factors}
\label{sec:factors}
The rational functions that occur in physical calculations usually include polynomial factors in the given set of variables. Once these factors are found, they lead to  simpler results and can be removed to reduce the number of probes required by black-box interpolations. Recently, \citeres{Smirnov:2020quc,Usovitsch:2020jrk} have shown that in the context of \abbrev{IBP} reductions denominators of rational functions can factorize completely by choosing a specific basis of master integrals.

In this section, we describe an algorithm implemented in \code{FireFly} that searches for univariate polynomial factors of black-box functions automatically as an additional scan before the actual interpolation.
These factors are then canceled in the interpolation to reduce the number of black-box probes and to simplify the results.
Factors including two or more variables are not covered by our algorithm. This restriction allows us to utilize the \code{FLINT} library\,\cite{flint,Hart:2010} for polynomial factorizations over finite fields. The factors in $z_i$ are calculated as follows:
\begin{enumerate}
\item Replace the variable set $\{z_1,\dots,z_n\}\setminus z_i$ by distinct random numbers $\{r_{j,1}\}\in\zp{p_1}$.
\item Evaluate the black box with distinct, randomized values $t$
at $\vec{z}=\{r_{1,1},\dots,r_{i-1,1},\allowbreak t,r_{i+1,1},\dots,r_{n,1}\}$ and perform a univariate rational-function interpolation to obtain a rational function in $z_i$ labeled as $f(z_i;\{r_{j,1}\};p_1)$.
\item Calculate the factors of numerator and denominator of $f(z_i;\{r_{j,1}\};p_1)$ in $z_i$.
\item Generate a new set of random numbers $\{r_{j,2}\}\in\zp{p_1}$ and perform a second interpolation to obtain the rational function $f(z_i;\{r_{j,2}\};p_1)$. Instead of performing an interpolation with Thiele's formula, one can build a system of equations because the monomial degrees are known from the first interpolation.
\item Calculate the factors of numerator and denominator of $f(z_i;\{r_{j,2}\};p_1)$ in $z_i$.
\item Compare the factors of $f(z_i;\{r_{j,1}\};p_1)$ and $f(z_i;\{r_{j,2}\};p_1)$ and accept only the coinciding factors.
Bring these common factors to canonical form in order to reconstruct their monomial coefficients.
The canonical polynomials are labeled as $q_{\mathrm{n}/\mathrm{d}}(z_i)$ for numerator (n) and denominator (d), respectively.
\item Repeat steps 1--6 over additional prime fields $\zp{p_k}$ and apply the Chinese remainder theorem\,\cite{von_zur_Gathen} to combine the results until the rational reconstruction\,\cite{Wang:1981,Monagan:2004} of the monomial coefficients of $q_{\mathrm{n}/\mathrm{d}}$ succeeds.
\end{enumerate}
Note that there is a chance to randomly choose $\{r_{j,l}\}$ such that terms vanish.
This probability is provided by the DeMillo-Lipton-Schwartz-Zippel lemma\,\cite{DeMillo:1978,Zippel:1979,Schwartz:1980} and can be reduced by increasing the characteristic of the prime field.

While scanning for factors in $z_i$, we also determine its maximum degree. After all $z_i$ are scanned, we perform an internal variable reordering with respect to the maximum degrees. The variable with maximum degree is set to one and only obtained via homogenization.

After a full scan, the obtained factors are used to cancel the corresponding terms of each black-box function.
They are included again in the final result.
Before returning the result to the user, the internal variable order is changed back to the user input.

To illustrate our algorithm, we perform the factor scan for the black-box function
\begin{equation}
f(z_1,z_2) = (z_1-1)(z_2-2)(z_1z_2-1)= -2+2z_1+z_2+z_1z_2-2z_1^2z_2-z_1z_2^2+z_1^2z_2^2 \,.
\label{eq:fac_ex}
\end{equation}
We start by scanning in $z_1$, while setting the first prime field to $\zp{17}$. The random number for $z_2$ is chosen to $z_2=r_{2,1}=5$.
Interpolating the dependence of $z_1$ with distinct and randomized $t$ and factorizing yields
\begin{equation}
f(z_1;\{5\};17) = 3+16z_1+15z_1^2 = 15(10+z_1)(16+z_1)\,.
\label{eq:fac_1}
\end{equation}
We now proceed by choosing $z_2=r_{2,2}=10$ which leads to
\begin{equation}
f(z_1;\{10\};17) = 8+14z_1+12z_1^2=12(5+z_1)(16+z_1)\,.
\label{eq:fac_2}
\end{equation}
By comparing \eqn{eq:fac_1} and \eqn{eq:fac_2}, we identify the common factor
$16+z_1$ which is independent of the choice of $z_2$. Therefore, we assume
that this term is a correct factor in $z_1$, which can be further processed by
rational-reconstruction algorithms. Indeed, applying the same procedure over a
different prime field and combining these results using the Chinese remainder
theorem as well as rational reconstruction results in the factor
\begin{equation}
q_\mathrm{n}(z_1) = z_1-1 \,.
\end{equation}
Analogously, we can find the factor $z_2-2$ in $z_2$ to arrive at the total factor
\begin{equation}
q_\mathrm{n}(z_1,z_2) = (z_1-1)(z_2-2) \,.
\end{equation}
As mentioned before, this algorithm cannot find the mixed factor $z_1z_2-1$ in \eqn{eq:fac_ex}. However, the remaining interpolation of $f(z_1,z_2)$ includes only this mixed term, when evaluating the black box as
\begin{equation}
\frac{f(y_1,y_2)}{q_\mathrm{n}(y_1,y_2)}
\end{equation}
at the values $y_1$ and $y_2$.

In addition to the algorithm above, one could also apply a different approach
by reconstructing rational functions in each variable univariately and only
perform the factorization over $\mathbb{Q}$. Also this reconstruction has to
be performed at least twice for random variable choices to identify the
factors that are independent of the other variables. However, since the interpolations are performed over fields with 63-bit characteristics, large coefficients can arise that require many prime fields to apply the rational reconstruction successfully, which leads to avoidable black-box probes. Furthermore, the factorization, which has to be computed after both reconstructions succeeded, takes additional computational effort as the numbers are not restricted to machine-size integers. Therefore, we do not provide an implementation of this alternative approach.

To quantify the improvements by this additional factor scan, we define the benchmark functions
{\allowdisplaybreaks
\begin{align}
f_1(z_1,\dots,z_4) &= \frac{\left(1+z_1+z_2+z_3+z_4\right)^{20}-1}{\left(1+z_1+z_2+z_3+z_4\right)^{20}-1+z_1^{20}} \,,
\label{eq:fac_b_1}
\\
f_2(z_1,\dots,z_4) &= (z_4-4)(z_4-3)\frac{\left(1+z_1+z_2+z_3+z_4\right)^{20}-1}{\left(1+z_1+z_2+z_3+z_4\right)^{20}-1+z_1^{20}} \,,
\label{eq:fac_b_2}
\\
f_3(z_1,\dots,z_{20}) &= \frac{\sum_{i=1}^{20}z_i^{20}}{\sum_{i=1}^{5}\left(z_1z_2+z_3z_4+z_5z_6\right)^i} \,,
\label{eq:fac_b_3}
\\
f_4(z_1,\dots,z_{20}) &= \frac{\sum_{i=1}^{20}z_i^{20}}{\sum_{i=1}^{5}\left(z_1z_2+z_3z_4+z_5z_6\right)^iz_{20}^{35}} \,,
\label{eq:fac_b_4}
\\
f_5(z_1,\dots,z_{20}) &= \frac{\sum_{i=1}^{20}z_i^{20}}{\sum_{i=1}^{5}\left(z_1z_2+z_3z_4+z_5z_6\right)^i(z_2z_{20}^{35}-1)} \,.
\label{eq:fac_b_5}
\end{align}}%
The function $f_1$ is almost completely dense and no univariate factor is present. $f_2$ is the same as $f_1$ with two additional factors in $z_4$. Factors of this type occur in many physical calculations with dimensional regularization. $f_3$, $f_4$, and $f_5$ are sparse functions with 20 variables. In $f_4$ the degree 35 term in $z_{20}$ in the denominator factorizes and in $f_5$ a variable reordering ($z_1\leftrightarrow z_{20}$) can simplify the interpolation.
We compare the interpolation of the benchmark functions without and with the factor scan in \tab{tab:fac_scan_results}.
The scan for a sparse shift is enabled by default.
\begin{table}[ht]
\begin{center}
\caption{Comparison of runtime and required probes for the interpolation and rational reconstruction of the benchmark functions defined by \eqs{eq:fac_b_1}--(\ref{eq:fac_b_5}) without and with the factor scan using a single thread of an \abbrev{AMD} Ryzen 5 3400G processor.}
\label{tab:fac_scan_results}
\begin{tabular}{c|c c|c c c}
\toprule
\multicolumn{1}{c|}{} & \multicolumn{2}{c|}{No factor scan} & \multicolumn{3}{c}{With factor scan}\\
Function & Probes & Runtime & Probes & Probes for scan & Runtime \\
\midrule
$f_1$ & 42627 & 3.6\,s & 42966 & 339 & 3.7\,s \\
$f_2$ & 50479 & 4.7\,s & 43060 & 433 & 3.7\,s \\
$f_3$ & 1615 & 0.04\,s & 2530 & 915 & 0.23\,s \\
$f_4$ & 38313 & 0.70\,s & 2563 & 949 & 0.23\,s \\
$f_5$ & 38523 & 0.72\,s & 3957 & 949 & 0.28\,s
\end{tabular}
\end{center}
\end{table}

For the dense function $f_1$, our algorithm does not find any factor. Consequently, there is no improvement. However, the additional probes required to search for factors are negligible compared to the total number of probes. Additionally, the runtime is only affected mildly.
When factors can be found, as is the case for $f_2$, the interpolation can succeed faster and with fewer probes, while only requiring about 100 additional probes for the scan compared to $f_1$. The increased amount of probes during the scan can be explained by the higher degree in $z_4$.
For sparse functions as $f_3$ and $f_4$, a possible improvement due to factorization depends strongly on the structure of the function. Since no factors in $f_3$ can be found by our algorithm, there is no improvement and the required probes increase by roughly $50\,\%$ compared to the interpolation without the scan. However, when high-degree factors can be found as in $f_4$, the number of black-box probes can decrease by an order of magnitude.
Although no factors can be found by our algorithm for $f_5$, reordering the variables leads to a significant improvement in the number of black-box probes by an order of magnitude. Since our shift-finder algorithm prefers variable shifts in the last variables, which are assumed to be of lower degree than the first variables, fewer additional monomials are generated by the shift, which further reduces the number of required black-box probes\,\cite{Klappert:2019emp}.

In our examples, the time for the black-box evaluations is negligible compared to the time for the interpolation. For physical applications, however, the evaluation time of the black box is usually dominant.
Therefore, we consider a smaller number of probes as more important than the additional computational work required to search for factors. The impact of our factorization algorithm in a physical context is illustrated by an \abbrev{IBP} reduction in \sct{sec:reduction}.

Also the benchmark functions defined by \eqs{eq:fac_b_1}--(\ref{eq:fac_b_5}) as well as an executable are available in the source code of \code{FireFly} since version \code{2.0.3}.
We refer to the file \code{README.md} accompanying the source code for instructions.

\section{Technical improvements and new features}
\label{sec:Technical_Improvements}
In this section we describe technical improvements and new features implemented in \code{FireFly} \code{2.0}. 
First, we changed the interface of \code{FireFly} according to the \emph{curiously recurring template pattern} (\abbrev{CRTP}), which allows the user to compute probes with arrays of fixed size in order to reduce the overhead for some problems.
The new interface is described in \sct{sec:interface}.
In \sct{sec:parser}, we describe the new shunting-yard parser to read in
expressions from files and strings in a way optimized for fast evaluations.
Lastly, we implemented support for the parallelization with \code{MPI}.
Our implementation and short instructions are described in \sct{sec:mpi}.

\subsection{New interface and overhead reduction}
\label{sec:interface}
\code{FireFly\;2.0} introduces some changes in the black-box interface. The black-box functor has been extended to template arguments following the \abbrev{CRTP}. This change allows for a static interface that can be used to evaluate the black box at sets of parameter points of different size.
This new design can reduce a potential overhead of the black box by performing operations over vectors instead of scalars. To build the black box, one has to define a black-box functor, whose evaluation member function accepts a template argument, i.e.
\begin{lstlisting}
class BlackBoxUser : public BlackBoxBase<BlackBoxUser> {
public:
  // Constructor
  BlackBoxUser(){};
  // Evaluation member function at the template argument FFIntTemp
  template<typename FFIntTemp>
  std::vector<FFIntTemp> operator()(const std::vector<FFIntTemp>& values) {
    // Define what the black box should do
    ...
  }
  // User-defined member functions
  ...
}
\end{lstlisting}
There are two changes in the interface shown in the above listing compared to the previous versions of \code{FireFly}. On the one hand, the definition of the functor is derived from its base class and takes itself as a template argument.
This is the realization of the \abbrev{CRTP} that allows for a static interface. On the other hand, the operator \code{()}, which is used to probe the black box at a given set of values is now templated. In practice, nothing changes for the user when these two modifications are implemented since the black box is only evaluated with objects that behave exactly like an \code{FFInt} as in previous versions of \code{FireFly}.
These new objects are implemented in the form of the \code{FFIntVec} classes, which are fixed size arrays of \code{FFInt} objects.
Currently, the sizes $2$, $4$, $8$, $16$, $32$, $64$, and $128$ are implemented.
Our implementation does not utilize the vectorization units of modern \abbrev{CPU}s.

By default, the black box is only evaluated at a single parameter point and returns the evaluations of all functions in the black box at this parameter point.
However, for some applications, like solving systems of equations, runtime improvements can be achieved when the system is solved once over an array of parameter points instead of several times over the individual parameter points.
Therefore, the \code{Reconstructor} class can now be initialized with a maximum \emph{bunch size} as
\begin{lstlisting}
Reconstructor<BlackBoxUser> reconst(n_vars,
                                    n_threads,
                                    maximum_bunch_size,
                                    black_box);
\end{lstlisting}
When the third argument is greater than one, the \code{Reconstructor} class will call the evaluate function of the black-box functor with \code{FFIntVec} objects, i.e.\ collections of parameter points, up to the maximum bunch size.
Each time a thread is assigned to compute a probe, it will gather probes to bunches
and will then compute the probes over this bunch in a single evaluation.
The size of the next bunch $b$ is calculated by the formula
\begin{equation}
  b = \min(2^{p_2}, b_\mathrm{max}) \quad\mathrm{with}\quad p_2 = \max\left(0, \left\lfloor\log_2\left(\frac{l_q}{t}\right)\right\rfloor\right)
\end{equation}
for $t$ threads, the maximum bunch size $b_\mathrm{max}$, and $l_q$ probes in the queue.
This design assures that a maximum amount of parallelism is favored over the reduction of potential overheads.
Using large bunch sizes usually increases the required amount of memory significantly, e.g.\ each entry in a system of equation contains bunch size \code{FFInt} objects instead of one.

The parser described in \sct{sec:parser} supports the evaluation in bunches and profits from it.

\subsection{Parsing collections of rational functions}
\label{sec:parser}
The first release version of \code{FireFly} expected the user to provide its implementation of the black box merely in native \code{C++} code. To simplify the usage of \code{FireFly}'s interpolation algorithms, we implemented a parser, which reads in text files or strings of formulas and rewrites them into an optimized format for further evaluation. These formulas can then be used, for example, as black-box functions, or to create a new black box based on these expressions.

Only rational functions with rational numbers as coefficients are supported.
They do not have to be provided in canonical form.
The accepted notation of parsable functions is inspired by the syntax of \code{Mathematica}\,\cite{mathematica}. A file can contain several functions separated by semicolons \code{;}, which mark the end of each function.
Variables can be composed of numbers and letters, i.e.\ elements of the set $\{\code{a,b,...,z,A,B,...,Z,0,1,2,...,9}\}$, with the restriction that a variable has to begin with a letter, e.g.\ \code{s12}. In addition, a variable is restricted to 16 characters. The supported binary arithmetic operators are
\begin{equation*}
\code{+}\,,\quad \code{-}\,,\quad \code{*},\quad \code{/}\,,\quad \code{\^{}}\,,
\end{equation*}
where \code{*} indicates a multiplication and \code{\^{}} an exponentiation operation. Moreover, the unary operators \code{+} and \code{-} are supported and \code{+-} or \code{-+} are interpreted as \code{-}. Implicit multiplication via the space character as well as any other combination of unary and binary operators are not supported. That is particularly important for negative exponents, e.g.\ \code{x\^{}(-10)}, that are only interpreted correctly when set in parentheses.

To illustrate the usage of the parser, the file \code{fun} with the content
\begin{lstlisting}
(-12 + 13/7*z1 + Z2^3 - (z1*Z2)^2)/z1;
\end{lstlisting}
shall be parsed.
This can be done by including the header file \code{ShuntingYardParser.hpp} of the \code{FireFly} library and the following line of code:
\begin{lstlisting}
firefly::ShuntingYardParser parser("fun", {"z1", "Z2"});
\end{lstlisting}
The \code{ShuntingYardParser} object is constructed with a path to a file that includes a collection of rational functions as the first argument and a vector of the occurring variables as strings as the second argument. During the construction of \code{parser}, the file \code{fun} is read in line by line and the string representing a function is joined up to the separator \code{;}. This string is then passed to an algorithm that rewrites the function to reverse Polish notation, also known as postfix notation, i.e.\ the function in the file \code{fun} becomes
\begin{equation*}
\code{-12 13 7 / z1 * + Z2 3 \^{} + z1 Z2 * 2 \^{} - z1 /}
\end{equation*}
and is stored inside \code{parser} as a vector of strings. The elements of this vector are referred to as tokens.
A function in postfix notation can be evaluated in linear time with respect to the amount of tokens using a \code{stack}.
The functions stored in \code{parser} can be evaluated at the values \code{y1} and \code{y2} by calling
\begin{lstlisting}
auto res = parser.evaluate({y1, y2});
\end{lstlisting}
Note that \code{y1} and \code{y2} can be \code{FFInt} or \code{FFIntVec} objects. \code{res} is a vector of \code{FFInt} or \code{FFIntVec} objects depending on the choice of \code{y1} and \code{y2}, where each entry in \code{res} corresponds to the evaluation of a function found in the file \code{fun}. The position of the evaluated function corresponds to the position in the file. As \code{fun} only contains one function, \code{res} is a vector of size one in the example.

Instead of constructing the \code{ShuntingYardParser} object with a file, one can also pass a vector of functions in string format as the first argument. For example
\begin{lstlisting}
firefly::ShuntingYardParser parser({{"z1+Z2"}, {"2*z1"}}, {"z1", "Z2"});
\end{lstlisting}
will construct \code{parser} with the two functions $z_1+Z_2$ and $2z_1$.

When a collection of functions is expected to be evaluated several times over the same prime field, the \code{ShuntingYardParser} class provides the optimized version \code{evaluate\_pre} of the member function \code{evaluate}, which makes use of pre-evaluated values for the tokens in the postfix notation. In practice, for the current prime field the images in $\zp{p}$ of the coefficients defined in $\mathbb{Q}$ are calculated only once and stored for later use. Hence, rational numbers like \code{13/7} are only evaluated once in $\zp{p}$. Thus, it is advisable to store rational numbers as such, e.g.\ \code{13/7*x} instead of \code{13*x/7}. Operators and variables are translated to an internal representation that is mapped to an integer value. These optimizations can lead to a significant performance improvement since the vector of strings is replaced by a vector of integers which is already evaluated except for the explicit values of the variables. To use this feature for the example above, one has to call: 
\begin{lstlisting}
parser.precompute_tokens();
auto res = parser.evaluate_pre({y1, y2});
\end{lstlisting}
The member function \code{precompute\_tokens} has to be called only once any time the underlying prime field changes. Calling the constructor of a \code{ShuntingYardParser} object automatically executes \code{precompute\_tokes}. \code{evaluate\_pre} returns the same object as \code{evaluate}.

In order to demonstrate the performance of our \code{ShuntingYardParser}, we generate dense multivariate polynomials of three variables up to a degree bound $D$ with monomial coefficients that are random rational numbers in the interval $[10^{98},10^{100}]$.\footnote{We provide a \code{Mathematica} script that generates such polynomials in \app{sec:poly_script}.} We also compare the performance without and with pre-evaluation.
The results are shown in \tab{tab:sy_parser_results}.
\begin{table}[htb]
\begin{center}
\caption{Comparison of runtime and memory consumption of the \code{ShuntingYardParser} object without and with pre-evaluated tokens. These benchmarks were run on a single thread of an \abbrev{AMD} Ryzen 5 3400G processor.}
\label{tab:sy_parser_results}
\begin{tabular}{l|c c |c c |c}
\toprule
\multicolumn{1}{c|}{} & \multicolumn{2}{c|}{No pre-evaluation} & \multicolumn{2}{c|}{With pre-evaluation} & \multicolumn{1}{c}{}\\
Function &  Parsing & Evaluation & Parsing & Evaluation & Memory \\
\midrule
$D=92$  & 0.6\,s & 0.28\,s & 0.8\;s & 0.02\,s & 0.25\,GiB \\
$D=159$  & 2.5\,s & 1.4\,s & 3.9\;s & 0.11\,s & 1.2\,GiB \\
$D=200$  & 5.3\,s & 3.0\,s & 7.7\;s & 0.22\,s & 2.4\,GiB \\
$D=252$  & 11\,s & 6.0\,s & 16\;s & 0.45\,s & 4.8\,GiB \\
\end{tabular}
\end{center}
\end{table}

Our reference polynomial is of maximum degree $D=92$ and, thus, consists of 138415 monomials.
The file size of this polynomial is 30\,MiB. The polynomials up to $D=159$, $D=200$, and $D=252$ are chosen such that the number of monomials is roughly a factor 5, 10, and 20, respectively, larger than the reference polynomial to demonstrate the scaling of the parser. The file sizes scale correspondingly. As expected, the parsing and evaluation times scale linearly in the number of monomials or, equivalently, in the number of operations. Although the parsing is up to 50\,\% slower when pre-evaluating values over a prime field, as one has to evaluate the function once, the evaluation time is an order of magnitude faster compared to no pre-evaluation.

Due to the linear scaling in the number of operations, polynomial representations that minimize this number, e.g.\ the Horner form, which is optimal in the number of operations, can lead to an additional improvement in the evaluation time.
Therefore, the impact of the Horner representation of a polynomial is shown in \tab{tab:sy_parser_results_h} for the same set of functions in comparison.
\begin{table}[htb]
\begin{center}
\caption{Comparison of runtime and memory consumption of the \code{ShuntingYardParser} object without and with pre-evaluated tokens using the Horner form. These benchmarks were run on a single thread of an \abbrev{AMD} Ryzen 5 3400G processor.}
\label{tab:sy_parser_results_h}
\begin{tabular}{l|c c |c c|c}
\toprule
\multicolumn{1}{c|}{} & \multicolumn{2}{c|}{No pre-evaluation} & \multicolumn{2}{c|}{With pre-evaluation} & \multicolumn{1}{c}{}\\
Function & Parsing & Evaluation & Parsing & Evaluation & Memory \\
\midrule
$D=92$ & 0.4\,s & 0.21\,s & 0.6\;s & 3.7\,ms & 0.18\,GiB \\
$D=159$ & 2.2\,s & 1.1\,s & 3.0\;s & 17\,ms & 0.8\,GiB\\
$D=200$ & 4.5\,s & 2.1\,s & 5.9\;s & 32\,ms & 1.6\,GiB \\
$D=252$ & 9.0\,s & 4.2\,s & 12\;s & 62\,ms & 3.1\,GiB \\
\end{tabular}
\end{center}
\end{table}

Utilizing this polynomial representation, the evaluation times can be decreased by an order of magnitude when using the pre-evaluation.
Without pre-evaluation, there are still small performance gains.
Hence, it is preferable to choose polynomial representations that minimize certain types of operations before evaluating them. Note that \code{FireFly} does not perform any optimization of polynomial representations except the pre-evaluation of tokens.

When parsing collections of functions, it is sometimes useful to omit redundant functions that might be evaluated several times. Especially during \abbrev{IBP} reductions, many coefficients in the system of equations are the same and only have to be evaluated once. For such cases, the \code{ShuntingYardParser} provides the additional option which checks the parsed functions for duplicates. This check can be enabled when constructing a parser object with a third, but optional, argument:
\begin{lstlisting}
firefly::ShuntingYardParser parser("fun", {"z1", "Z2"}, true);
\end{lstlisting}
By setting the third argument to \code{true}, the \code{ShuntingYardParser} evaluates the parsed functions for different randomized sets of variables and prime fields and searches for equal evaluations. Duplicate functions are removed and replaced by a key. Therefore, no function is evaluated more than once during a single evaluation step. The returned vector of the evaluation member functions remains unchanged. By default, no check for redundant functions is performed.

The parser also supports the evaluations in bunches described in \sct{sec:interface}.
\tab{tab:sy_parser_bunches} shows the interpolation and reconstruction of the polynomial with $D=200$ in Horner form using different bunch sizes.
The \abbrev{CPU} time for each probe drops by about 30\,\% and the total runtime still by 20\,\% when increasing the bunch size from 1 to 128.
Since this is not a memory intensive problem after reading in the file, the memory consumption stays almost the same for all bunch sizes.
\begin{table}[ht]
\begin{center}
\caption{Reconstruction of the polynomial with $D=200$ in Horner form using different maximal bunch sizes.
These benchmarks were performed on cluster nodes equipped with two Intel Xeon Platinum 8160 processors with 24 cores each and hyperthreading disabled.}
\label{tab:sy_parser_bunches}
\begin{tabular}{c|c c c}
\toprule
Bunch size & Runtime per probe & Total runtime \\
\midrule
1 & 50\,ms & 6\,h 30\,min \\
2 & 47\,ms & 6\,h 10\,min \\
4 & 43\,ms & 5\,h 50\,min \\
8 & 40\,ms & 5\,h 40\,min \\
16 & 40\,ms & 5\,h 40\,min \\
32 & 37\,ms & 5\,h 20\,min \\
64 & 36\,ms & 5\,h 20\,min \\
128 & 34\,ms & 5\,h 10\,min \\
\end{tabular}
\end{center}
\end{table}

Lastly, \code{FireFly}'s parser is compatible with \code{Mathematica}'s \code{InputForm} stored to a file.

\subsection{\code{MPI}}
\label{sec:mpi}
In addition to the parallelization with \code{C++} threads, \code{FireFly} now supports parallelization with \code{MPI}, which enables the user to utilize multiple nodes on a computer cluster.
We employ a master and worker pattern.
Both master and worker processes use their own thread pools for the internal parallelization of each process and the communication between the individual processes is handled by the \code{MPI} protocol.

The master process runs the \code{Reconstructor} class as it would in the case without \code{MPI}, i.e.\ it handles the interpolation and reconstruction of the rational functions.
If threads are idle, it also computes probes.
The worker processes run the \code{MPIWorker} class which only computes probes.
Each process has to provide a single thread to handle the communication with \code{MPI}.
The master process distributes the required probes onto all processes available, including itself, and the worker processes send the results back when they are finished.

To use \code{MPI} an \code{MPI} library like \code{Open\,MPI}\,\cite{openmpi,Gabriel:2004} or \code{MPICH}\,\cite{mpich} has to be installed.
We advise the latter for optimal performance.
Additionally, \code{FireFly} has to be compiled with the option \code{-DWITH\_MPI=true}.

The basic code structure for the user is the following.
First, the \code{MPI} environment has to be initialized.
The thread support of \code{MPI} has to be set to \code{MPI\_THREAD\_SERIALIZED} or \code{MPI\_THREAD\_MULTIPLE}, because multiple threads can make \code{MPI} calls.
However, we made sure that only one thread can make calls at a time.
Thus, we advise to use the former for better performance.
Then, the node has to find out whether it should run as the master or a worker process.
In both cases, one has to define the black box.
For the master process one creates the \code{Reconstructor} object and runs it exactly as in the case without \code{MPI}.
For the worker processes, the user has to create \code{MPIWorker} objects.
They are initialized similarly to the \code{Reconstructor} object and require the number of variables, the number of threads to use, the maximum bunch size, and the black box as arguments.
The initialization of the \code{MPIWorker} objects already runs them.
If the calculation of the \code{Reconstructor} on the master process finishes, they receive a signal to shut down.
Finally, one has to finish the \code{MPI} environment.
\begin{lstlisting}
// Initialization of MPI processes
int provided;
MPI_Init_thread(NULL, NULL, MPI_THREAD_SERIALIZED, &provided);

int process;
MPI_Comm_rank(MPI_COMM_WORLD, &process);

// Create the user-defined black box
BlackBoxUser black_box();

if (process == firefly::master) {
  firefly::Reconstructor<BlackBoxUser> reconst(n_vars,
                                               n_threads,
                                               maximum_bunch_size,
                                               black_box);

  reconst.reconstruct();
} else {
  firefly::MPIWorker<BlackBoxUser>(n_vars,
                                   n_threads,
                                   maximum_bunch_size,
                                   black_box);
}

// Finish MPI environment
MPI_Finalize();
\end{lstlisting}

We provide the file \code{example\_mpi.cpp} as example.
It is automatically compiled together with \code{FireFly} when setting \code{-DWITH\_MPI=true}.
After compiling, one can run the code with \code{MPI} by calling
\begin{lstlisting}
mpiexec -n $NUMBER_OF_PROCESSES $EXECUTABLE
\end{lstlisting}
Since each \code{MPI} process has to allocate one thread for communication, one should only run one \code{MPI} process on each node and assign all threads of this node to it.
For example, if two nodes with 48 threads each are available, running two \code{MPI} processes with 48 threads each yields a better performance than running four \code{MPI} processes with 24 threads each.
For more command-line options of \code{MPI}, we refer to the respective manuals.

We again use the polynomial with $D=200$ in Horner form from \sct{sec:parser} to illustrate the performance gain by using several nodes of a computer cluster.
\tab{tab:sy_parser_mpi} shows the scaling with 1, 2, and 4 nodes with 48 cores each.
Using 2 nodes decreases the runtime by 26\,\% compared to a single node.
However, doubling the nodes again only decreases it by 16\,\%.
Thus, the scaling is not linear with the number of nodes.
This is expected because the interpolation itself is performed on a single thread and is quite expensive for this dense high degree function.
Only the computation of the probes utilizes all available threads.
Moreover, they are very cheap individually for this synthetic benchmark as shown in \sct{sec:parser}.
The performance gain is much higher for problems with a large number of expensive probes.
One notable example are \abbrev{IBP} reductions (s.\ \sct{sec:IBPs}) and we refer to \citere{Klappert:2020nbg} for the application of \code{MPI} in this context.
\begin{table}[ht]
\begin{center}
\caption{Reconstruction of the polynomial with $D=200$ in Horner form from \sct{sec:parser} using \code{MPI} on cluster nodes equipped with two Intel Xeon Platinum 8160 processors with 24 cores each and hyperthreading disabled.}
\label{tab:sy_parser_mpi}
\begin{tabular}{c|c c c c}
\toprule
Nodes & 1 & 2 & 4 \\
\midrule
Total runtime & 6\,h 30\,min & 4\,h 40\,min & 3\,h 50\,min
\end{tabular}
\end{center}
\end{table}

\section{Applications in \abbrev{IBP} reductions}
\label{sec:IBPs}
During multi-loop calculations in high-energy physics one usually encounters large sums of $L$-loop scalar Feynman integrals
\begin{equation}
  \label{eq:scalar_int}
  I(d, \{p_j\}, \{m_i\}, \{a_i\}) \equiv \int_{k_1,\dots ,k_L} \frac{1}{P_1^{a_1}\dots P_N^{a_N}}
\end{equation}
with
\begin{equation}
  \int_k \equiv \int \frac{\mathrm{d}^d k}{(2\pi)^d}
\end{equation}
and the inverse propagators, i.e.\ $P_i = q_i^2 - m_i^2 + \mathrm{i}\epsilon$ in Minkowski space.
The $q_i$ are linear combinations of the loop momenta $k_l$ and the external momenta $p_j$.
The integral $I(d, \{p_j\}, \{m_i\}, \{a_i\})$ depends on the space-time dimension $d$, the set of masses $\{m_i\}$, the set of external momenta $\{p_j\}$, and the propagator powers $a_i$ which take integer values.
The number of propagators $N$ is restricted by $L$ and the number of external momenta $E$ to
\begin{equation}
  N = EL + \frac{L(L+1)}{2} \,.
\end{equation}
It is useful to define the sum of all positive powers of the propagators of an integral as
\begin{equation}
  r \equiv \sum_{i=1}^{N} \theta \left(a_i -\frac{1}{2}\right) a_i
\end{equation}
and the absolute value of the sum of all negative powers as
\begin{equation}
  s \equiv \sum_{i=1}^{N} \theta \left(\frac{1}{2}- a_i\right) |a_i| \,,
  \label{eq:def_s}
\end{equation}
where $\theta(x)$ is the Heaviside step function.
Usually, an integral with higher $r$ or higher $s$ is regarded more difficult than an integral with lower $r$ or $s$.
Therefore, $r$ and $s$ can be used to sort the occurring integrals by difficulty.

Many of the scalar Feynman integrals given by \eqn{eq:scalar_int} can be solved by the integration-by-parts (\abbrev{IBP}) strategy of Chetyrkin and Tkachov\,\cite{Tkachov:1981wb,Chetyrkin:1981qh}.
They observed that inserting the scalar product of a derivative with respect to a loop momentum with another momentum into \eqn{eq:scalar_int} leads to a vanishing integral in dimensional regularization:
\begin{equation}
  \int_{k_1,\dots k_L} \frac{\partial}{\partial k_i^\mu} \left( \tilde q_j^\mu \frac{1}{P_1^{a_1}\dots P_N^{a_N}} \right) = 0 \,,
\end{equation}
where $\tilde q_j^\mu$ can either be another loop momentum or an external momentum.
By explicitly evaluating the derivative one arrives at the linear relations
\begin{equation}
  \label{eq:IBP}
  0 = \sum_n c_n I(d, \{p_j\}, \{m_i\}, \{a_i^{(n)}\})
\end{equation}
with modified $a_i^{(n)}$, where the values change by the addition or subtraction of small integers.
The coefficients $c_n$ are rational functions in $d$, $\{m_i\}$, and $\{p_j\cdot p_k\}$ with a small degree and also depend on the $a_i$ in general.
These relations are called \abbrev{IBP} relations.

The most prominent strategy utilizing \abbrev{IBP} relations is the Laporta algorithm\,\cite{Laporta:2001dd}:
By inserting integer values for the $a_i$ in \eqn{eq:IBP}, one obtains a system of equations for the integrals.
This system can be solved to express the integrals through a basis of master integrals, which have to be computed by other methods, see e.g.\ \citere{Smirnov}.
The Laporta algorithm has been implemented in the public codes \code{AIR}\,\cite{Anastasiou:2004vj}, \code{FIRE}\,\cite{Smirnov:2008iw,Smirnov:2013dia,Smirnov:2014hma,Smirnov:2019qkx}, \code{Reduze}\,\cite{Studerus:2009ye,vonManteuffel:2012np}, and \code{Kira}\,\cite{Maierhoefer:2017hyi,Maierhofer:2018gpa}.
However, the systems of equations for state-of-the-art calculations become huge and expensive to solve both in terms of memory and runtime, amongst other things due to large intermediate expressions.

Since the \abbrev{IBP} relations are linear, the solution strategies only involve the addition, subtraction, multiplication, and division of the coefficients $c_n$.
Therefore, the coefficients of the master integrals are also rational functions in $d$, $\{m_i\}$, and $\{p_j\cdot p_k\}$.
Thus, the problems of the Laporta algorithm can be eased by finite-field techniques\,\cite{Kauers:2008zz}.
One can replace all occurring variables by elements of a finite field and solve the system of equations numerically, which is in general orders of magnitude faster than solving the system analytically.
This can be used to identify and remove the linearly dependent equations before the actual analytic reduction\,\cite{Kant:2013vta}.
However, one can also replace the analytic solution by numerical solutions altogether by employing interpolation techniques\,\cite{vonManteuffel:2014ixa}.
The first public code implementing this strategy is \code{FIRE6}\,\cite{Smirnov:2019qkx}, even though at least one private code exists for quite some time\,\cite{vonManteuffel:2016xki}.
Based on these techniques, also new strategies for \abbrev{IBP} reductions are pursued\,\cite{Bendle:2019csk,Guan:2019bcx}.
For the related strategy based on generalized unitarity, the usage of more advanced techniques from computer science has been pioneered in \citere{Peraro:2016wsq}.

We combined the integral-reduction program \code{Kira}\,\cite{Maierhoefer:2017hyi,Maierhofer:2018gpa} with \code{FireFly} to utilize finite-field and interpolation techniques for \abbrev{IBP} reductions and the improvements in \code{FireFly\;2.0}.
This new version of \code{Kira} is published with the separate publication\,\cite{Klappert:2020nbg}, which describes details of the implementation and provides sophisticated benchmarks.
Hence, we restrict ourselves to a single example in \sct{sec:reduction}, which nonetheless makes the performance improvements in \code{FireFly\;2.0} evident.

The calculation of physical observables usually requires the insertion of the reduction table into a sum of Feynman integrals given by \eqn{eq:scalar_int}.
For example, this has to be done to compute amplitudes as intermediate step of cross-section calculations.
In these insertion steps, a problem similar to the reduction arises, because both the table and the expression can become huge for state-of-the-art problems.
However, the final result expressed in terms of a sum of master integrals usually contains simpler coefficients compared to intermediate steps.
Therefore, these problems can also be eased with the help of interpolation techniques over finite fields.
The optimal strategy would be to interpolate the final coefficients without generating analytic reduction tables as intermediate step.
Admittedly, these tables are still required for some strategies to compute the master integrals\,\cite{Smirnov}.
Thus, as a first step we developed a tool which can insert reduction tables into expressions using \code{FireFly}.
It is described in \sct{sec:insertIBPs}.

\subsection{\abbrev{IBP} reduction with \code{Kira} and \code{FireFly}}
\label{sec:reduction}
\begin{figure}[ht]
\centering\includegraphics[trim = 0 1.5mm 0 1mm]{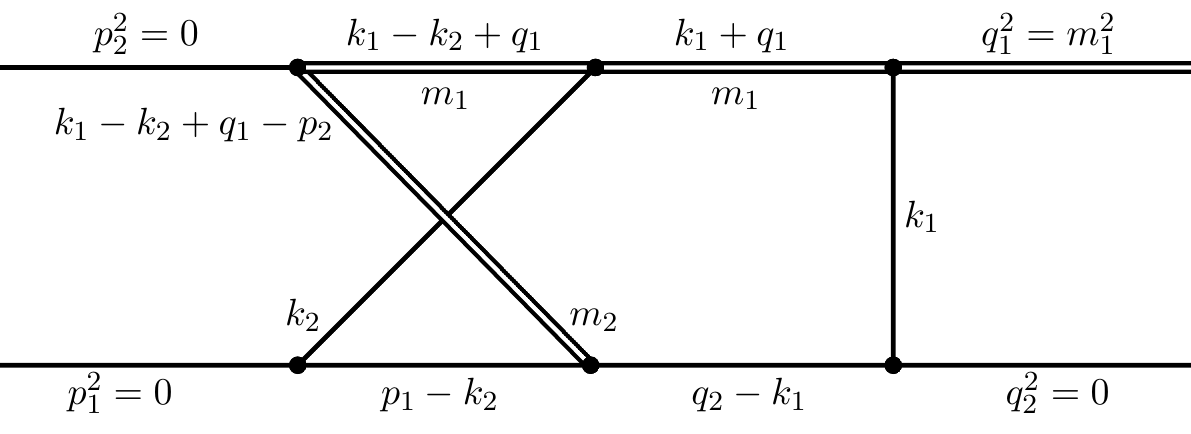}
\caption{The non-planar double box \code{topo5} which occurs, e.g., in single top production. The diagram was produced with \code{FeynGame}\,\cite{Harlander:2020cyh}.}
\label{fig:topo5}
\end{figure}

To illustrate the performance improvements of \code{FireFly\;2.0} in physical problems, we perform the \abbrev{IBP} reduction of the topology \code{topo5} shown in \fig{fig:topo5}.
We choose the physical value $r_\text{max} = 7$, set the scale $m_1 = 1$, and use the option \code{select\_mandatory\_recursively} with the same range of $r_\text{max}$ and $s_\text{max}$.
We compare a development version of \code{Kira} both without and with \code{FireFly\;2.0}.
For the algebraic reduction, this version of \code{Kira} contains only bug fixes and no performance improvements over \code{Kira\;1.2}.
In addition, we also include an older version of \code{Kira} which runs with \code{FireFly\;1.0}.
All of the versions are compiled with the \code{g++} compiler in version \code{9.2.0}\,\cite{gcc} and the standard memory allocator is replaced by linking \code{jemalloc} in version \code{5.2.1}\,\cite{jemalloc} to the executables.
Using an improved memory allocator leads to significant performance improvements for memory intensive computations like solving a system of equations on many threads.
Both versions of \code{FireFly} use \code{FLINT}\,\cite{flint,Hart:2010} for the modular arithmetic.
The algebraic version of \code{Kira} performs most of the computations with help of \code{Fermat\;6.31}\,\cite{fermat}.
These examples are solved on cluster nodes equipped with two Intel Xeon Platinum 8160 processors with 24 cores each and 192\,GiB of \abbrev{RAM} in total.
Hyperthreading is disabled.

The results are shown in \tab{tab:topo5}.
\begin{table}[ht]
  \begin{center}
    \caption{Comparison of runtime and memory consumption for \code{Kira}, \code{Kira} with \code{FireFly\;1.0}, and \code{Kira} with \code{FireFly\;2.0} in the reduction of \code{topo5} with the configuration described in the main text.}
    \label{tab:topo5}
    {\renewcommand{\arraystretch}{1.3}
    \begin{tabular}{c|c c|c c|c c}
      \toprule
      & \multicolumn{2}{c|}{\code{Kira}} & \multicolumn{2}{c|}{\code{Kira + FireFly\,1.0}} & \multicolumn{2}{c}{\code{Kira + FireFly\,2.0}} \\
      $s_\text{max}$ & Runtime & Memory & Runtime & Memory & Runtime & Memory \\
      \midrule
      1 & 2\,min 20\,s & 6.7\,GiB & 1\,min 30\,s & 1.1\,GiB & 46\,s & 1.3\,GiB \\
      2 & 2\,h & 28\,GiB & 1\,h & 2.5\,GiB & 17\,min & 3.1\,GiB \\
      3 & 15\,h 30\,min & 70\,GiB & 9\,h & 13\,GiB & 2\,h 30\,min & 11\,GiB \\
      4 & 92\,h & 164\,GiB & 90\,h & 56\,GiB & 20\,h & 45\,GiB \\
      \bottomrule
    \end{tabular}}
  \end{center}
\end{table}
Note that these numbers cannot directly be compared to the numbers in \citere{Klappert:2019emp}, because different machines and memory allocators are used.
The performance with \code{FireFly\;2.0} increases by a factor of 3--4 compared to \code{FireFly\;1.0}.
For higher $s_\text{max}$, even the memory consumption decreases slightly due to the scan for univariate factors as discussed below.

\tab{tab:topo5_details} shows a detailed comparison of \code{Kira} with \code{FireFly\;1.0} and \code{Kira} with \code{FireFly\;2.0}.
\begin{table}[ht]
  \begin{center}
    \caption{Detailed comparison of \code{Kira} with \code{FireFly\;1.0} and \code{Kira} with \code{FireFly\;2.0} in the reduction of \code{topo5} with the configuration described in the main text.}
    \label{tab:topo5_details}
    {\renewcommand{\arraystretch}{1.3}
    \begin{tabular}{c|c c c|c c c}
      \toprule
      & \multicolumn{3}{c|}{\code{Kira + FireFly\,1.0}} & \multicolumn{3}{c}{\code{Kira + FireFly\,2.0}} \\
      $s_\text{max}$ & Probes & \begin{tabular}{@{}c@{}}\abbrev{CPU} time \\ per probe\end{tabular} & \begin{tabular}{@{}c@{}}\abbrev{CPU} time \\ for probes\end{tabular} & Probes & \begin{tabular}{@{}c@{}}\abbrev{CPU} time \\ per probe\end{tabular} & \begin{tabular}{@{}c@{}}\abbrev{CPU} time \\ for probes\end{tabular} \\
      \midrule
      1 & 15300 & 0.16\,s & 97\,\% & 5200 & 0.14\,s & 95\,\% \\
      2 & 330100 & 0.43\,s & 98\,\% & 107500 & 0.40\,s & 97\,\% \\
      3 & 1127000 & 0.95\,s & 92\,\% & 433300 & 0.90\,s & 96\,\% \\
      4 & 4304500 & 1.9\,s & 81\,\% & 1624200 & 1.9\,s & 96\,\% \\
      \bottomrule
    \end{tabular}}
  \end{center}
\end{table}
The gain in runtime shown in \tab{tab:topo5} can be mainly attributed to the lower number of probes required, which decreased roughly by a factor of three throughout the whole range of $s_\text{max}$.
In addition, each probe became cheaper, because we replaced the parser of \code{pyRed} by our shunting-yard parser for which we use the pre-evaluation and removal of redundant functions as described in \sct{sec:parser}.
With \code{FireFly\;2.0}, the forward elimination dominates the whole process with about 94--95\,\%, whereas the initiation contributes with 3--4\,\% and the back substitution with only 1--2\,\%.
Even though the new interpolation algorithms described in \sct{sec:algorithmic_improvements} are more complicated than the ones in \code{FireFly\;1.0}, the \abbrev{CPU} time for the probes still completely dominates the reduction.
Furthermore, some performance improvements in \code{FireFly} even increase the percentage for the probes.

The impact of the scan for univariate factors is shown in \tab{tab:topo5_scan}.
\begin{table}[ht]
  \begin{center}
    \caption{Impact of the factor scan on the reduction of \code{topo5} with \code{Kira} in combination with \code{FireFly\;2.0}.
    We use the configuration described in the main text.}
    \label{tab:topo5_scan}
    {\renewcommand{\arraystretch}{1.3}
    \begin{tabular}{c|c c c|c c c}
      \toprule
      & \multicolumn{3}{c|}{No factor scan} & \multicolumn{3}{c}{With factor scan} \\
      $s_\text{max}$ & Runtime & Memory & Probes & Runtime & Memory & Probes \\
      \midrule
      1 & 58\,s & 1.3\,GiB & 9000 & 46\,s & 1.3\,GiB & 5200 \\
      2 & 25\,min & 3.4\,GiB & 174000 & 17\,min & 3.1\,GiB & 107500 \\
      3 & 3\,h 50\,min & 15\,GiB & 703700 & 2\,h 30\,min & 11\,GiB & 433300 \\
      4 & 33\,h & 67\,GiB & 2745200 & 20\,h & 45\,GiB & 1624200 \\
      \bottomrule
    \end{tabular}}
  \end{center}
\end{table}
It reduces the number of probes and with it the runtime by roughly 40\,\%.
In addition, it requires less memory especially for the more complicated reductions with higher $s_\text{max}$.
The main reason for these improvements is the fact that only one variable has to be shifted to generate constant terms for all coefficients.
In contrast, three variables have to be shifted without the scan and, thus, more artificial terms have to be interpolated and stored.
Without the scan, the memory consumption increases by more than 10\,\% compared to \code{FireFly\;1.0}.

\subsection{Insertion of \abbrev{IBP} tables}
\label{sec:insertIBPs}
The replacement tables obtained by \abbrev{IBP} reductions are usually huge
for state-of-the-art calculations. With increasing complexity of these
results, utilizing algebraic tools like \code{Mathematica}\,\cite{mathematica}
or \code{Fermat}\,\cite{fermat} can become unfeasible in order to perform simplifications during the insertion into expressions like amplitudes.
A more efficient approach can be to insert the \abbrev{IBP} tables numerically and
perform an interpolation of the final result over finite fields, e.g.\ with
\code{FireFly}. In order to automatize the insertion step and to illustrate
applications of interpolation algorithms, \code{FireFly\;2.0} provides the
additional executable \code{ff\_insert}, which is compiled by
default. \code{ff\_insert} can be found in the build directory or, when
installed, in the \code{bin} directory inside the installation directory. Its
usage requires the user to provide the directories \code{config} and \code{replacements} containing all required files.
The directory structure is as follows:
\begin{lstlisting}
config/
  |__ vars
  |__ functions
  |__ skip_functions
replacements/
  |__ $REPLACEMENT_LIST_1
  |__ ...
\end{lstlisting}
The \code{vars} and \code{functions} files contain the list of occurring
variables and the list of integral families, respectively, separated by new lines. In the \code{skip\_functions} file one can specify master integrals that should be omitted during the insertion. In the \code{replacements} directory, several files with arbitrary file name can be stored containing replacement rules for integrals in \code{Mathematica} syntax.
\code{FireFly} expects the formulas in a format, where the integral appears first followed by a potential coefficient, e.g.,
\begin{equation*}
\code{F1[1,0,1,-2]*(s+t+d)/42} + \code{F1[1,1,1,1]*(d-3)}\,,
\end{equation*}
where \code{F1} is an integral family defined in \code{functions} and \code{s}, \code{t}, and \code{d} are variables defined in \code{vars}.

The insertion of the replacement rules into the expression \code{\$INPUT\_FILE} can be started by calling
\begin{lstlisting}
ff_insert $INPUT
\end{lstlisting}
in the directory, where the \code{config} and \code{replacements} directories
can be found. The last command line option (\code{\$INPUT}) should be an input file, e.g.\ an
amplitude, or a directory containing several expressions for which replacements should be performed. The expressions have to be provided in the same syntax as the replacement tables. The following options can be specified:
\begin{lstlisting}
-p $NUMBER_OF_THREADS   // Defines the number of threads
-bs $MAXIMUM_BUNCH_SIZE // Defines the maximum bunch size
-m                      // Merge expressions in the given directory
-nfs                    // Disables the factor scan
-ni                     // Stores the unsimplified coefficients
-s                      // Enables saving of states
-h                      // Show options
\end{lstlisting}

\code{FireFly} will read in the expression in \code{\$INPUT} first.
Afterwards, the replacement tables are parsed and only the required replacement rules are selected. \code{FireFly} does not support repeated replacements. They can be obtained by either preparing the tables or by reusing the \code{ff\_insert} executable multiple times. Afterwards, the replacements are performed and a list of remaining integrals is created, which are assumed to be master integrals. Each coefficient of a master integral is passed to \code{FireFly}'s parser and then used for the interpolation. The functional interpolations are performed sequentially, i.e.\ each coefficient of a master integral is interpolated independently of the others. When all interpolations succeeded, the file \code{out\_\$INPUT} is written to the current directory. It contains a sum of master integrals and their respective coefficients. When only performing simplifications of master-integral coefficients, the \code{replacements} directory is not mandatory.

Since the interpolation of each master integral coefficient can be performed irrespective of the others, one can in principle use separate machines for each function. Therefore, the option \code{-ni} indicates that no interpolation has to be performed. Instead, all unsimplified coefficients are written to files in the newly created \code{coefficients} directory. They can be used as input for the \code{ff\_insert} executable or other processing.

We provide example data for an insertion job to be run with the \code{ff\_insert} executable in the \code{examples} directory of the \code{FireFly} files.

\section{Conclusions}
We described the main improvements implemented into \code{FireFly} since the release last year.
On the algorithmic side, we implemented the racing algorithm of \citeres{Kaltofen:2000,Kaltofen:2003} to improve the univariate interpolation of sparse polynomials as sub-algorithm for the interpolation of rational functions.
Moreover, we described the hybrid racer algorithm for rational functions in \sct{sec:hybrid_racer}, which removes the oversampling for dense functions compared to the algorithm of \citere{Cuyt:2011}.
Since the results of physical problems often contain simple factors as shown in \citeres{Smirnov:2020quc,Usovitsch:2020jrk}, we implemented an additional scan to search for univariate factors to remove them in the actual interpolation.

On the technical side, we changed the interface of \code{FireFly} to allow for a reduction of overhead by computing several probes in a single evaluation.
We also implemented a shunting-yard parser for expressions stored in files and
strings, which can be used to construct black boxes.
Moreover, \code{FireFly} now supports \code{MPI} in order to utilize computer clusters.

The impact of the algorithmic improvements on a physical \abbrev{IBP} reduction has been studied in \sct{sec:reduction} by combining the program \code{Kira} with \code{FireFly}.
In this context, we also provide a tool based on \code{FireFly}, which allows to insert replacement tables into expressions.
The new version of \code{Kira} supporting reductions with \code{FireFly} is publicly available accompanied by the separate publication\,\cite{Klappert:2020nbg}.

\code{FireFly\;2.0} is publicly available at
\begin{center}
  \href{https://gitlab.com/firefly-library/firefly}{\code{https://gitlab.com/firefly-library/firefly}} \,.
\end{center}

\section*{Acknowledgments}
We thank Philipp Maierh\"ofer for the idea of overhead reductions with bunches and discussions about this topic.
We thank Johann Usovitsch for inspiring the factor scan and discussions about this topic.
Furthermore, we would like to thank Long Chen, Robert Harlander, Philipp Maierh\"ofer, and Johann Usovitsch for comments on the manuscript.
We thank Herschel Chawdhry, Long Chen, Wen Chen, Joshua Davies, Jerry Dormans, Philipp Maierh\"ofer, Mario Prausa, Robert Schabinger, Johann Usovitsch, and Alexander Voigt for bug reports and feature suggestions.

This research was supported by the \textit{Deutsche Forschungsgemeinschaft} (\abbrev{DFG}, German Research Foundation) under grant \href{http://gepris.dfg.de/gepris/projekt/396021762?language=en}{396021762} -- \href{http://p3h.particle.kit.edu/start}{TRR 257}.
F.L.\ acknowledges financial support by \abbrev{DFG} through project \href{http://gepris.dfg.de/gepris/projekt/386986591?language=en}{386986591}.
Some of the calculations were performed with computing resources granted by RWTH Aachen University under project rwth0541.

\clearpage
\appendix

\section{\code{Mathematica} code to generate dense polynomials}
\label{sec:poly_script}
In this appendix we provide \code{Mathematica} functions that generate random
polynomials in either Horner or expanded form. They read
\begin{lstlisting}
polygenHorner[{x_,y_,z_}, degree_]:= HornerForm[Sum[
    RandomInteger[{1,100}]/RandomInteger[{1,100}]*x^i*y^j*z^k
    , {i, 0, degree}, {j, 0, degree - i}, {k, 0, degree - i - j}
  ]];
polygen[{x_,y_,z_}, degree_]:= Sum[
    RandomInteger[{1,100}]/RandomInteger[{1,100}]*x^i*y^j*z^k
    , {i, 0, degree}, {j, 0, degree - i}, {k, 0, degree - i - j}
  ];
\end{lstlisting}
and generate polynomials of three variables, \code{x}, \code{y}, and
\code{z}, up to a given degree. When performing the replacements
\begin{lstlisting}
rep1 = {Rational[a_,b_] -> Rational[RandomInteger[{a*10^98,a*10^99}],
    RandomInteger[{b*10^98,b*10^99}]]};
rep2 = {Plus[a_Integer,b__] -> Plus[
    Rational[RandomInteger[{a*10^98,a*10^99}],
    RandomInteger[{3*a*10^98, 3*a*10^99}]],b]};
\end{lstlisting}
the coefficients of each monomial are randomized rational numbers in the range $[10^{98},10^{100}]$. The obtained polynomials which are not in Horner form are then rewritten to a format, where the rational number is in front of the monomial.

\clearpage

\IfFileExists{./\jobname_ref.tex}{
  
}{}

\end{document}